\documentclass[prl,english,reprint,nofootinbib,aps,superscriptaddress,showpacs,showkeys]{revtex4-1}

\usepackage{babel,calc,amsmath,amsthm,amssymb,graphicx,subfigure,xcolor,mathtools,
      longtable,braket,bm,extarrows}

\usepackage[T1]{fontenc}
\usepackage[unicode=true]{hyperref}
\hypersetup{
     colorlinks=true,       		% false: boxed links; true: colored links
     linkcolor=blue,          	% color of internal links
     citecolor=red,            % color of links to bibliography
     urlcolor=magenta,           	% color of external links
}
\newtheorem*{theorem*}{Theorem}

\newtheorem*{corollary*}{Corollary}

\newtheorem*{lemma*}{Lemma}

\newtheorem*{proposition*}{Proposition}

\newtheorem*{conjecture*}{Conjecture}
\theoremstyle{definition}

\newtheorem*{definition*}{Definition}
\theoremstyle{remark}

\newtheorem*{remark*}{Remark}
\begin{document}
\renewcommand{\figureautorefname}{Fig.}
\renewcommand{\figurename}{Fig.}
\renewcommand{\tableautorefname}{Tab.}
\renewcommand{\tablename}{Tab.}

%\title{Optimal $N$-Setting Linear Einstein-Podolsky-Rosen Steering Inequalities}

%\title{Optimal Einstein-Podolsky-Rosen Steering Inequalities for Werner State}

\title{Revealing the Boundary between Quantum Mechanics and Classical Model by EPR-Steering Inequality}

\author{Ruo-Chen Wang}
\affiliation{School of Physics, Nankai University, Tianjin 300071, People's Republic of China}

\author{Zhuo-Chen Li}
\affiliation{School of Physics, Nankai University, Tianjin 300071, People's Republic of China}

\author{Xing-Yan Fan}
\affiliation{Theoretical Physics Division, Chern Institute of Mathematics, Nankai
      University, Tianjin 300071, People's Republic of China}

\author{Xiang-Ru Xie}
\affiliation{School of Physics, Nankai University, Tianjin 300071, People's Republic of China}

\author{Hong-Hao Wei}
\affiliation{School of Physics, Nankai University, Tianjin 300071, People's Republic of China}

\author{Choo Hiap Oh}
\email{phyohch@nus.edu.sg}
\affiliation{Centre for Quantum Technologies and Department of Physics, National University of Singapore, 117543, Singapore}

\author{Jing-Ling Chen}
\email{chenjl@nankai.edu.cn}
\affiliation{Theoretical Physics Division, Chern Institute of Mathematics, Nankai
      University, Tianjin 300071, People's Republic of China}

\date{\today}

	\begin{abstract}
In quantum information, the Werner state is a benchmark to test the boundary between quantum mechanics and classical models. There have been three well-known critical values for the two-qubit Werner state, i.e., $V_{\rm c}^{\rm E}=1/3$ characterizing the boundary between entanglement and separable model, $V_{\rm c}^{\rm B}=1/K_G(3)$ characterizing the boundary between Bell's nonlocality and the local-hidden-variable model, %$\big(K_G(3)$ expresses the Grothendieck constant of order three$\big)$, 
while $V_{\rm c}^{\rm S}=1/2$ characterizing the boundary between Einstein-Podolsky-Rosen (EPR) steering and the local-hidden-state model. So far, the problem of $V_{\rm c}^{\rm E}=1/3$ has been completely solved by an inequality involving in the positive-partial-transpose criterion, while how to reveal the other two critical values by the inequality approach are still open. In this work, we focus on EPR steering, which is a form of quantum nonlocality intermediate between entanglement and Bell's nonlocality. By proposing the optimal $N$-setting linear EPR-steering inequalities, we have successfully obtained the desired value $V_{\rm c}^{\rm S}=1/2$ for the two-qubit Werner state, thus resolving the long-standing problem.
	\end{abstract}
	

	\maketitle

	%\tableofcontents
	
	%\section{Introduction}
		
    \emph{Introduction.---}In 1935, Einstein, Podolsky and Rosen (EPR) published a milestone work by proposing the famous EPR paradox for an entangled continuous-variable state~\cite{EPR}. In the same year, from EPR's paper Schr\"odinger extracted two fundamental concepts of quantum nonlocality, namely, ``quantum entanglement'' and ``EPR steering''~\cite{Schrodinger35}. Bell made an important step forward in 1964 by considering a version based on the entanglement of two spin-1/2 particles introduced by Bohm, and then presented the third fundamental concept (i.e., ``Bell's nonlocality'') through quantum violation of Bell's inequality for entangled states~\cite{Bell}. In 2007, Wiseman, Jones, and Doherty presented a unified treatment for quantum nonlocality from the viewpoint of quantum information tasks~\cite{WJD07,WJD07PRA}, thus triggering a wave of research on EPR steering.

\begin{figure}[t]
		\centering
		\includegraphics[width=80mm]{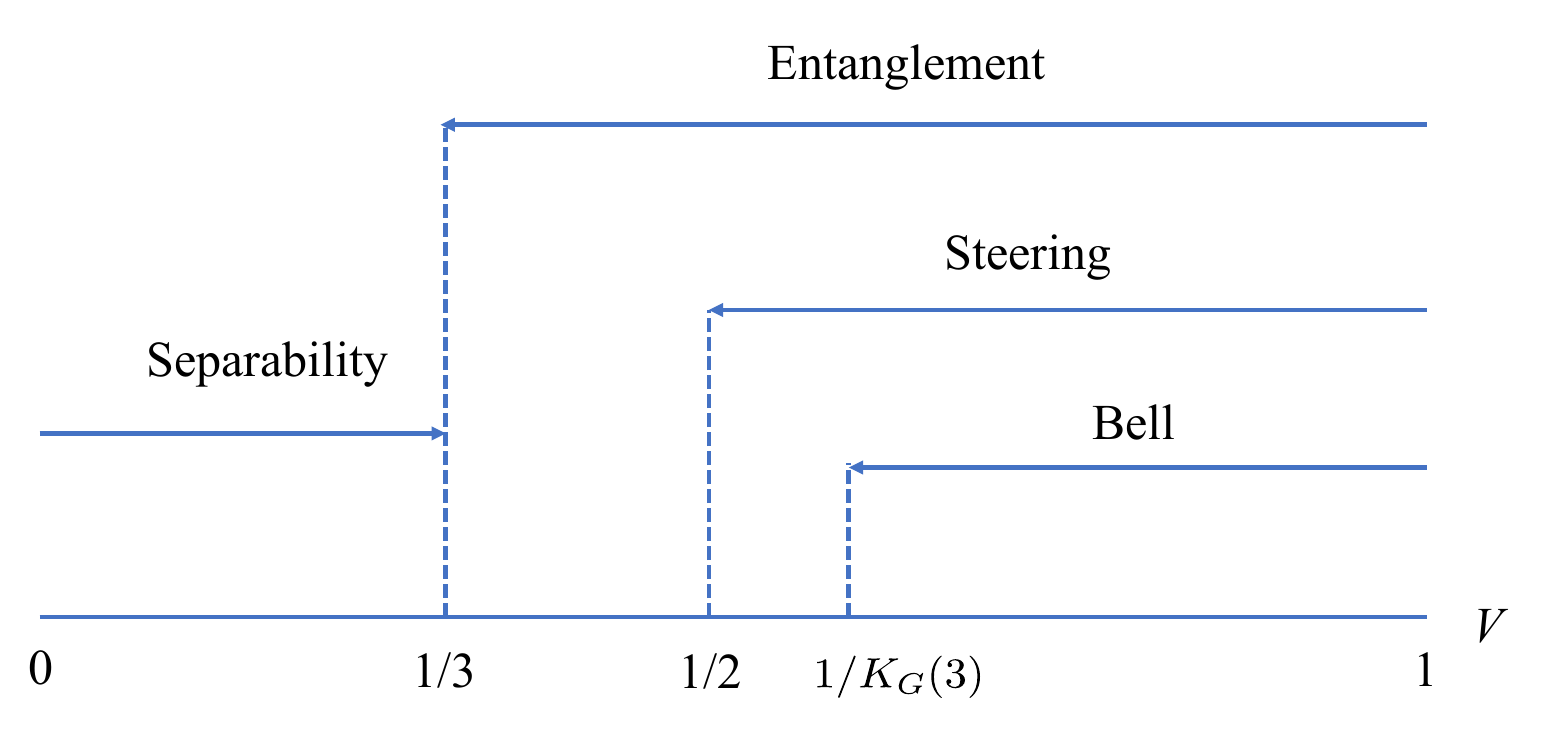}
		\caption{Illustration of the hierarchical structure of quantum nonlocality. Let us take the two-qubit Werner state as an example. The critical values for entanglement, steering, and Bell's nonlocality are $V_{\rm c}=1/3$, $1/2$, and $1/K_G(3)$, respectively. Thus quantum states that have Bell's nonlocality form a subset of EPR steerable states, which in turn form a subset of entangled states.}\label{fig:CritPoint}
	\end{figure}

    According to Wiseman \emph{et al.}'s work, quantum nonlocality was classified into three distinct types: entanglement, steering, and Bell's nonlocality, and they possessed a hierarchical structure. Explicitly, Bell's nonlocality is the strongest-type nonlocality in nature, quantum entanglement is the weakest-type nonlocality, while EPR steering is a novel form of quantum nonlocality intermediating between them. For instance, the hierarchical structure can be demonstrated through the well-known two-qubit Werner state \cite{werner89}, which is given by
    \begin{eqnarray}
		&& \rho_{\rm W} = V |{\Phi_-}\rangle\langle{\Phi_-}|  + (1-V) \rho_{\rm noise},
	\end{eqnarray}
    where
   % \begin{eqnarray}
%		&& |{\Phi_-}\rangle= \frac{1}{\sqrt{2}}(|{01}\rangle-|{10}\rangle)
%	\end{eqnarray}
    $|{\Phi_-}\rangle= (|{01}\rangle-|{10}\rangle)/\sqrt{2}$ is the maximally entangled state, $\rho_{\rm noise}=\mathbb{I}/4$ is the white noise (i.e., the maximally mixed state), $\mathbb{I}$ is the $4\times 4$ identity matrix, $V$ is a parameter characterizing the mixture degree of the two-qubit maximally entangled state with the white-noise density matrix, and $V\in[0, 1]$. There are two extremes in $\rho_{\rm W}$, so it is not difficult to imagine that, when $V$ slides from $1$ to $0$, it will cross the boundary between quantum mechanics and classical models. Until now there have been three well-known critical values for the two-qubit Werner state, namely, $V_{\rm c}^{\rm E}=1/3$, $V_{\rm c}^{\rm S}=1/2$, and $V_{\rm c}^{\rm B} =1/K_G(3)$ for quantum entanglement, EPR steering, and Bell's nonlocality, respectively~\cite{2006PRAcin,CASA2011} (see Fig. \ref{fig:CritPoint}). Here $K_G(3)$ indicates the Grothendieck constant of order three \cite{2017QHirsch}, and according to Ref. \cite{2023PRRDesignolle}, $K_G(3)\in(1.4367,1.4546)$.

    Usually, quantum nonlocality is defined through its corresponding classical model. For example, entanglement is defined by its counterpart, i.e., the separable model; if a state cannot be described by the separable model, then the state is entangled~\cite{werner89,RMP2009}. Similarly, the corresponding classical models for Bell's nonlocality and EPR steering are the local-hidden-variable (LHV) model~\cite{RMP2014} and local-hidden-state (LHS) model~\cite{2017RPPCavalcanti,RMP2020,2022PRXQXiang,2024CPWang}, respectively. Therefore, it is straight forward to judge whether a quantum state has nonlocal property (or not) using the definition of quantum nonlocality. In 1989, Werner presented the strict definition of separable model and showed that the state $\rho_W$ does not have Bell's nonlocality but is still entangled in the region $V\in (1/3, 1/2]$, thus pointing out that quantum entanglement and Bell's nonlocality are two totally different concepts~\cite{werner89}. This is also the first time that the critical value $V_{\rm c}=1/2$ appearing in the literature. However, this critical value is not exactly the boundary between quantum entanglement and Bell's nonlocality. In 2007, by proposing the strict definition of LHS model, Wiseman \emph{et al.} pointed out that $V_{\rm c}=1/2$ is nothing but exactly the critical value of EPR steering for the two-qubit Werner state~\cite{WJD07}.

    However, judging the nonlocal property of quantum states directly by its definition is inefficient and inconvenient. Subsequently more efficient approaches to detect quantum nonlocality are developed. Traditionally there are two efficient approaches. One is the inequality approach, and the other is the approach of nonlocality without inequality, i.e., the all-versus-nothing (AVN) proof of nonlocality. For instance, for quantum entanglement, there are the positive-partial-transpose (PPT) criterion~\cite{1996PRAPPT} and other abundant entanglement witnesses~\cite{RMP2009}; for Bell's nonlocality, there are the approach of Bell's inequalities~\cite{1969CHSH,CGLMP}, and the approach of the Greenberger-Horne-Zeilinger paradox~\cite{GHZ89} as well as Hardy's paradox~\cite{Hardy93,Chen2018}. Similarly, people have presented the inequality approach for EPR steering, such as the linear EPR-steering inequality~\cite{NP2010}, chained EPR-steering inequalities~\cite{Meng2018a}, and so on~\cite{RMP2020}. Moreover, the AVN proofs of EPR steering have also been investigated~\cite{AVN2013,Chen2016}, which reveal the steering property of quantum states through logic contradictions.

    Revealing the boundary between quantum mechanics and classical models by the inequality approach is very significant in quantum foundations. For the first critical value $V_{\rm c}^{\rm E}=1/3$, it has been completely determined by the following inequality
     \begin{eqnarray}
		&& \lambda_{\rm min} \left(\varrho^{\rm PT} \right)\geq 0,
	\end{eqnarray}
    where $\varrho^{\rm PT}$ represents the partial transpose of the density matrix $\varrho$, and $\lambda_{\rm min}$ denotes its minimal eigenvalue.     By solving $\lambda_{\rm min} \left(\rho_{\rm W}^{\rm PT} \right)=0$, one may easily work out the critical value of $V=1/3$. In 2006, Ac{\' i}n \emph{et al.} pointed out that the critical value $V_{\rm c}^{\rm B}\approx 0.6595$ is the boundary between Bell's nonlocality and LHV model ~\cite{2006PRAcin,2009ManyQuest}. The approach of Bell's inequalities can be adopted to show Bell's nonlocality. For example, the famous Clauser-Horne-Shimony-Holt (CHSH) inequality has revealed that the two-qubit Werner state possesses Bell's nonlocality for $V>1/\sqrt{2}$~\cite{1969CHSH}. Later on, by developing more efficient Bell's Inequalities for Werner states,
    V\'ertesi~\cite{2008PRAVertesi}  and other researchers~\cite{2015JPAHua} have decreased the parameter $V$ to a lower value $0.69$. In principle, by exhausting all possible Bell's inequalities (including infinite-setting cases), one can achieve the critical value $V_{\rm c}^{\rm B}$. However, it is an impossible mission, and thus such a problem remains open.

    In this work, we focus on EPR steering. By studying the optimal $N$-setting linear EPR-steering inequalities, we successfully obtained the desired value $V_{\rm c}^{\rm S}=1/2$ for the two-qubit Werner state when $N\rightarrow \infty$, thus resolving the long-standing problem.

%\newpage
%
%    The purpose of this work is to advance the study of the linear EPR-steering inequality. The paper is organized as follows. In Sec. II, we briefly review the linear EPR-steering inequalities given in Ref.~\cite{NP2010}. In Sec. III, numerically  we present the optimal $N$-setting linear EPR-steering inequalities up to $N=18$. In Sec. IV, we discuss the $N$-setting linear EPR-steering inequality for the limit of $N\rightarrow \infty$, and obtain successfully the desired critical value $V_{\rm c}=1/2$ for the two-qubit Werner state. In Sec. V., some applications of the optimal EPR-steering inequalities are presented, which can be used to detect more quantum states. Conclusion is made in the last section.

%	\section{Brief Review of the Linear EPR-Steering Inequality}

  \emph{Optimal $N$-Setting Linear EPR-Steering Inequalities.}---In 2010, Saunders, Jones, Wiseman, and Pryde (SJWP) greatly promoted the problem of
  $V_{\rm c}^{\rm S}=1/2$ by presenting the $N$-setting linear EPR-steering inequalities (in this paper we call them as SJWP's inequalities, in order to distinguish from the optimal ones)~\cite{NP2010}. For a two-qubit system shared by Alice and Bob, the $N$-setting linear EPR-steering inequalities are given by
  \begin{eqnarray}\label{eq:ineq}
		\mathcal{I}_{N} = \frac{1}{N}\sum_{j=1}^{N}\langle A_j \hat{\sigma}_j^B\rangle \overset{{\rm{LHS}}}{\leq} \mathcal{C}_N.
	\end{eqnarray}
  Here $A_j$ is the $j$-th measurement result of Alice with  $A_j\in \{+1, -1\}$, $\hat{\sigma}_j^B\equiv \vec{b}_j \cdot \vec{\sigma}^B$, $\vec{b}_j$  is the $j$-th measurement direction of Bob, $\vec{\sigma}=(\sigma_x, \sigma_y, \sigma_z)$ is the vector of Pauli matrices, and $j=1, 2,..., N$. The parameter $\mathcal{C}_N$ describes the classical bound (i.e., the LHS bound), which is calculated through
  \begin{eqnarray}
		&&\mathcal{C}_N = \frac{1}{N}  \mathop{\rm max}\limits_{\{A_j\}}\left\{\lambda_{\rm max}\left[\sum_{j=1}^{N}A_j\hat{\sigma}^{B}_j\right]\right\},
	\end{eqnarray}
with $\lambda_{\rm max}[\mathcal{M}]$ being the maximal eigenvalue of matrix $\mathcal{M}$.

  %      \onecolumngrid
%      \begin{widetext}
      \begin{table}[h]
	\centering
\caption{The measurement directions and the corresponding LHS bounds for SJWP's inequalities.}
\begin{tabular}{llc}
\hline\hline
$N$ &  Measurement directions $\vec{b}_j$   & LHS bound $\mathcal{C}_N$   \\
  \hline
2 & $\vec{b}_1 = (1, 0, 0)$ & $\dfrac{\sqrt{2}}{2}\approx 0.7071$ \\
 & $\vec{b}_2 = (0, 1, 0)$ &  \\
  \hline
3 & $\vec{b}_1 = (1, 0, 0)$ & $\dfrac{\sqrt{3}}{3}\approx 0.5773$ \\
 & $\vec{b}_2 = (0, 1, 0)$ &  \\
  & $\vec{b}_3 = (0, 0, 1)$ &  \\
   \hline
4 & $\vec{b}_1 = \frac{1}{3}(\sqrt{6}, 0, \sqrt{3})$ & $\dfrac{\sqrt{3}}{3}\approx 0.5773$ \\
 & $\vec{b}_2 = \frac{1}{3}(-\sqrt{6}, 0, \sqrt{3})$ &  \\
  & $\vec{b}_3 = \frac{1}{3}(0, \sqrt{6},  \sqrt{3})$ &  \\
    & $\vec{b}_4 = \frac{1}{3}(0, -\sqrt{6}, \sqrt{3})$ &  \\
   \hline
6 & $\vec{b}_1 = (0, 0, 1)$ & $\dfrac{1+\sqrt{5}}{6}\approx 0.5393$ \\
 & $\vec{b}_2 = \left(\frac{2}{\sqrt{5}},0,\frac{1}{\sqrt{5}}\right)$ &  \\
  & $\vec{b}_3 = \left(\frac{\left(5-\sqrt{5}\right)}{10} ,\sqrt{\frac{1}{10} \left(\sqrt{5}+5\right)},\frac{1}{\sqrt{5}}\right)$ &  \\
    & $\vec{b}_4 = \left(-\frac{\left(5+\sqrt{5}\right)}{10} ,\sqrt{\frac{1}{10} \left(5-\sqrt{5}\right)},\frac{1}{\sqrt{5}}\right)$ &  \\
      & $\vec{b}_5 = \left(-\frac{\left(5+\sqrt{5}\right)}{10} ,-\sqrt{\frac{1}{10} \left(5-\sqrt{5}\right)},\frac{1}{\sqrt{5}}\right)$ &  \\
    & $\vec{b}_6 = \left(\frac{\left(5-\sqrt{5}\right)}{10} ,-\sqrt{\frac{1}{10} \left(\sqrt{5}+5\right)},\frac{1}{\sqrt{5}}\right)$ &  \\
   \hline
10 & $\vec{b}_1 = (0, 0, 1)$ & $\dfrac{3+\sqrt{5}}{10}\approx 0.5236$ \\
 & $\vec{b}_2 = \left(\frac{2}{3},0,\frac{\sqrt{5}}{3}\right)$ &  \\
  & $\vec{b}_3 = \left(\frac{\sqrt{5}}{3},\frac{1}{\sqrt{3}},\frac{1}{3}\right)$ &  \\
    & $\vec{b}_4 = \left(\frac{\left(3-\sqrt{5}\right)}{6} ,\sqrt{\frac{1}{6} \left(3+\sqrt{5}\right)},\frac{1}{3}\right)$ &  \\
      & $\vec{b}_5 = \left(-\frac{1}{3},\frac{1}{\sqrt{3}},\frac{\sqrt{5}}{3}\right)$ &  \\
    & $\vec{b}_6 = \left(\frac{\left(-\sqrt{5}-3\right)}{6} ,\sqrt{\frac{1}{6} \left(3-\sqrt{5}\right)},\frac{1}{3}\right)$ &  \\
  & $\vec{b}_7 = \left(\frac{\left(-\sqrt{5}-3\right)}{6} ,-\sqrt{\frac{1}{6} \left(3-\sqrt{5}\right)},\frac{1}{3}\right)$ &  \\
    & $\vec{b}_8 = \left(-\frac{1}{3},-\frac{1}{\sqrt{3}},\frac{\sqrt{5}}{3}\right)$ &  \\
      & $\vec{b}_9 = \left(\frac{\left(3-\sqrt{5}\right)}{6} ,-\sqrt{\frac{1}{6} \left(3+\sqrt{5}\right)},\frac{1}{3}\right)$ &  \\
    & $\vec{b}_{10} = \left(\frac{\sqrt{5}}{3},-\frac{1}{\sqrt{3}},\frac{1}{3}\right)$ &  \\
%  \hline
% & & \\
 \hline\hline
\end{tabular}\label{tab:NP}
\end{table}
%\end{widetext}
%\twocolumngrid

%\newpage

  Quantum mechanically, $A_j$ corresponds to the operator $\hat{A}_j\equiv \hat{\sigma}_j^A=\vec{a}_j \cdot \vec{\sigma}^A$, and $\vec{a}_j$  is the $j$-th measurement direction of Alice. Based on the expectation values $\langle \Phi_-| \hat{\sigma}_j^A \hat{\sigma}_j^B| \Phi_-\rangle
= - \hat{a}_j \cdot \hat{b}_j$
%  \begin{eqnarray}
%		 \langle \Phi_-| \hat{\sigma}_j^A \hat{\sigma}_j^B| \Phi_-\rangle
%&=& \langle \Phi_-| (\vec{a}_j \cdot \vec{\sigma}^A) \otimes (\vec{b}_j \cdot \vec{\sigma}^B) | \Phi_-\rangle\nonumber\\
%&=& - \hat{a}_j \cdot \hat{b}_j,
%	\end{eqnarray}
and ${\rm Tr}\left[\rho_{\rm noise} (\hat{\sigma}_j^A \hat{\sigma}_j^B)\right]=0$,
% \begin{eqnarray}
%		&& {\rm Tr}\left[\rho_{\rm noise} [(\vec{a}_j \cdot \vec{\sigma}^A) \otimes (\vec{b}_j \cdot \vec{\sigma}^B)]\right]=0,
%	\end{eqnarray}
one easily knows that, for the Werner state with a fixed $V$, the maximal quantum value of $\mathcal{I}_{N}$ (denoted by $\mathcal{Q}_N$) is always
 \begin{eqnarray}
		&& \mathcal{Q}_N=V,
	\end{eqnarray}
because Alice can always choose her measurement direction as $\vec{a}_j=-\vec{b}_j$ such that $\langle \Phi_-| \hat{\sigma}_j^A \hat{\sigma}_j^B| \Phi_-\rangle=1$ for any $j=1, 2, ..., N$. Therefore, for the inequality (\ref{eq:ineq}), it may detect the steering property of Werner state in the region $V\in (\mathcal{C}_N, 1]$.

Putting an explicit set of $N$ measurement directions $\vec{b}_j$ into (\ref{eq:ineq}), then one immediately has a concrete $N$-setting linear EPR-steering inequality. Ref.~\cite{NP2010} has provided a family of $N$-setting linear EPR-steering inequalities with $N=2, 3, 4, 6, 10$. In Table \ref{tab:NP}, the measurement directions and the corresponding LHS bounds for SJWP's inequalities have been listed. One may find that the value $\mathcal{C}_N$ decreases when $N$ grows. For $N=10$, the classical bound is $\mathcal{C}_{10}\approx 0.5236$, thus the $10$-setting SJWP's inequality can detect the steering property of $\rho_{\rm W}$ for the region $V\in (\mathcal{C}_{10}, 1]$. SJWP have designed the $N$ measurement directions
by using the geometric symmetry of Platonic solids. However, since the number of Platonic solids in three-dimensional space is limited, so they stopped at $N=10$. Nonetheless, $\mathcal{C}_{10}$ is very close to $0.5$, this fact hints that the $N$-setting linear EPR-steering inequality is a suitable candidate to solve the problem of $V_{\rm c}^{\rm S}=1/2$, provided we can design the appropriate measurement directions.

SJWP have not yet mentioned whether their inequalities are optimal~\cite{NP2010}. In this work, we come to study the optimal $N$-setting linear EPR-steering inequalities. Here ``optimal'' means that, for a fixed $N$, one can find out the $N$ optimal measurement directions $\vec{b}_j$ such that the LHS bound $\mathcal{C}_N$ in (\ref{eq:ineq}) is the lowest. Here we use the simulated annealing algorithm (assisted by numerical analysis in Mathematica) to compute the optimal value $\mathcal{C}_N$. This algorithm is an optimization algorithm commonly used to find optimal solutions in large search spaces. It is inspired by the metal annealing process and searches for the global optimal solution in the solution space by gradually reducing the temperature~\cite{2011OPTIMIZATION}.	

Numerically, we have successfully found the optimal linear EPR-steering inequalities up to $N=20$. The results of $N\leq 10$ are listed in Table \ref{tab:optimal1}, and those of $10 <N\leq 20$ are given in Supplemental Material (SM) \cite{SM}. For $N=2$ and $3$, our results are essentially the same (up to a rotation) as those in Table \ref{tab:NP}. This means that SJWP's inequalities are optimal for $N=2, 3$, but they are no longer optimal  for $N\geq 4$. For $N=4$ to $6$, we can actually obtain the analytical expressions for measurement directions and LHS bounds. While for $N>7$, we can only provide the numerical results for measurement directions and LHS bounds. When $N=20$, $\mathcal{C}_{20}\approx 0.5073$, which deviates from $V_{\rm c}^{\rm S}=1/2$ merely by $0.007$.

\emph{Optimal Inequality for Infinity Settings.}---Let us study the linear EPR-steering inequality for $N\rightarrow \infty$. There are some useful clues involved in Table \ref{tab:optimal1}. One may observe that: (i) After marking these measurement directions $\vec{b}_j$ in the unit sphere, one finds that these $N$ points are distributed on a hemisphere. Due to symmetry, it does not matter how to take the hemisphere  on the sphere. For simplicity, we have designed the hemisphere as the \emph{Northern Hemisphere}, i.e., the $z$-component of all vectors $\vec{b}_j$'s are non-negative; (ii) For some $\vec{b}_j$'s, their $z$-components are equal, e.g., for $N=6$, one finds that ${b}_1^z={b}_2^z={b}_3^z=1/\sqrt{10}$, ${b}_4^z={b}_5^z=2/\sqrt{10}$. This means that these three points $\vec{b}_1, \vec{b}_2, \vec{b}_3$ are located on a circle parallel to the equatorial plane, and another two points $\vec{b}_4, \vec{b}_5$ are located on another circle that is also parallel to the equatorial plane. This suggests us to cut the hemisphere into many circles	when considering $N\rightarrow \infty$; (iii) Let $\vec{u}$ be the summation of all $\vec{b}_j$'s divided by $N$, one then finds that $\vec{u}$ points to the north pole, whose magnitude is just equal to $\mathcal{C}_N$, i.e., $\vec{u}=\vec{e}_z\, \mathcal{C}_N$. This implies that the maximal value of $\mathcal{I}_N$ may occur at all $A_j$'s equal to $+1$, and the LHS bound can be calculated through $\mathcal{C}_N=|\vec{u}|$, thus simplifying the calculations.

\onecolumngrid
\begin{widetext}
\begin{table}[t]
	\centering
\caption{The measurement directions and the corresponding LHS bounds for optimal inequalities. Here $N=2$ to $10$.}
\begin{tabular}{llc}
\hline\hline
$N$\;\;\;\; &  Measurement directions $\vec{b}_j$   & \;\;LHS bound $\mathcal{C}_N$   \\
  \hline
%  & & \\
2 &$\boxed{\vec{b}_1 = (0, -\frac{1}{\sqrt{2}}, \frac{1}{\sqrt{2}}), \; \vec{b}_2 = (0, \frac{1}{\sqrt{2}}, \frac{1}{\sqrt{2}})}$ & $\frac{\sqrt{2}}{2}\approx 0.7071$
\vspace{1mm} \\
  \hline
3 & $\boxed{\vec{b}_1 = (\frac{\sqrt{6}}{3}, 0, \frac{\sqrt{3}}{3}), \; \vec{b}_2 = (-\frac{\sqrt{6}}{6}, \frac{\sqrt{2}}{2}, \frac{\sqrt{3}}{3}), \; \vec{b}_3 = (-\frac{\sqrt{6}}{6}, -\frac{\sqrt{2}}{2}, \frac{\sqrt{3}}{3})}$ & $\frac{\sqrt{3}}{3}\approx 0.5773$ \vspace{1mm}\\
   \hline
4 & $\boxed{\vec{b}_1 = (\frac{2}{\sqrt{5}},0, \frac{1}{\sqrt{5}}), \; \vec{b}_2 = (-\frac{1}{2\sqrt{5}}, \frac{\sqrt{3}}{2}, \frac{1}{\sqrt{5}}), \; \vec{b}_3 = (-\frac{7}{4\sqrt{5}}, -\frac{\sqrt{3}}{4}, \frac{1}{\sqrt{5}})}, \; \vec{b}_4 =  (\frac{1}{4\sqrt{5}}, -\frac{\sqrt{3}}{4}, \frac{2}{\sqrt{5}})$ & $\dfrac{\sqrt{5}}{4} \approx 0.5590$ \vspace{1mm}\\
   \hline
5 & $\boxed{\vec{b}_1 = (-0.9297 , 0 , 0.3683),\; \vec{b}_2 = (0.1459 , -0.9182 , 0.3683),\; \vec{b}_3 = (0.6837 , 0.6300 , 0.3683)},$ & $\dfrac{1}{20} \sqrt{2(9+\sqrt{33})}$ \vspace{1mm}\\
& $\vec{b}_4 = (-0.2460 , 0.6300 , 0.7366),\; \vec{b}_5 = (0.3461 , -0.3418 , 0.8737)$ &  $\approx 0.5430$  \\
\hline
6 & $\boxed{\vec{b}_1 = (\frac{3}{\sqrt{10}}, 0, \frac{1}{\sqrt{10}}),\; \vec{b}_2 = (-\frac{1}{3\sqrt{10}}, \frac{2\sqrt{2}}{3}, \frac{1}{\sqrt{10}}),\; \vec{b}_3 = (-\sqrt{\frac{2}{5}}, -\frac{1}{\sqrt{2}}, \frac{1}{\sqrt{10}})},$ & $\dfrac{\sqrt{10}}{6} \approx 0.5270$ \vspace{1mm}\\
    & $\boxed{\vec{b}_4 = (-\frac{7}{3\sqrt{10}}, \frac{1}{3\sqrt{2}}, \frac{2}{\sqrt{10}}),\; \vec{b}_5 = (\frac{1}{\sqrt{10}}, -\frac{1}{\sqrt{2}}, \frac{2}{\sqrt{10}})},\; \vec{b}_6 = (\frac{1}{3} \sqrt{\frac{2}{5}}, \frac{1}{3\sqrt{2}}, \frac{3}{\sqrt{10}})$ &  \vspace{1mm}\\
   \hline
7 & $\boxed{\vec{b}_1 = (0.8110, 0.5183, 0.2712),\; \vec{b}_2 = (-0.9625, 0, 0.2712)}, $ & $ 0.5268$ \vspace{1mm}\\
& $\vec{b}_3 = (-0.0256, -0.9423, 0.3337),\; \vec{b}_4 = (0.6536, -0.5082, 0.5609),$ &  \vspace{1mm}\\
& $\boxed{\vec{b}_5 = (-0.5502, 0.5183, 0.6547),\; \vec{b}_6 = (0.1845, 0.7330, 0.6547)},$ &  \vspace{1mm}\\
& $\vec{b}_7 = (-0.1108, -0.3192, 0.94126)$ & \\
\hline
8 & $\boxed{\vec{b}_1 = (0.9687, 0.0657, 0.2395),\; \vec{b}_2 = (-0.7471, -0.6200, 0.2395)},\; \vec{b}_3 = (-0.2099, 0.9422, 0.2611), $ & $ 0.5219$ \vspace{1mm}\\
    & $\boxed{\vec{b}_4 = (-0.6502, 0.4930, 0.5782),\; \vec{b}_5 = (0.6228, -0.5271, 0.5782),\; \vec{b}_6 = (-0.0879, -0.8111, 0.5782)},$ &  \vspace{1mm}\\
    & $\vec{b}_7 = (0.4317, 0.4921, 0.7560),\; \vec{b}_8 = (-0.3280, -0.0348, 0.9446)$ &  \\
   \hline
9 & $\boxed{\vec{b}_1 = (-0.9470, -0.2399, 0.2138),\;\vec{b}_2 = (0.9753, 0.0559, 0.2138)},$ & $0.5198$\vspace{1mm} \\
  & $\boxed{\vec{b}_3 = (0.2236, 0.9006, 0.3727),\;\vec{b}_4 = (-0.4840, 0.7917, 0.3727)},$ & \vspace{1mm}\\
  & $\boxed{\vec{b}_5 = (0.4918, -0.7506, 0.4412),\;\vec{b}_6 = (-0.2434, -0.8638, 0.4412)},$ & \vspace{1mm}\\
  & $\boxed{\vec{b}_7 = (0.5709, 0.1215, 0.8120),\;\vec{b}_8 = (-0.5810, -0.0558, 0.8120)},$ & \vspace{1mm}\\
  & $\vec{b}_9 = (-0.0062, 0.0404, 0.9992)$ & \\
  \hline
 10& $\boxed{\vec{b}_1 = (0.9160, 0.3469, 0.2016),\; \vec{b}_2 =(-0.9680, 0.1498, 0.2016)}, $ & $0.5168$ \vspace{1mm}\\
    & $\boxed{\vec{b}_3 =(-0.3817, -0.8738, 0.3013),\; \vec{b}_{4} = (0.5544, -0.7758, 0.30130)},$ &  \vspace{1mm}\\
    & $\boxed{\vec{b}_5 = (0.5288, 0.6830, 0.5039), \; \vec{b}_6 =(-0.6588, 0.5587, 0.5039)},\; \vec{b}_7 =(-0.0810, 0.7739, 0.6281),  $ &  \vspace{1mm}\\
        & $\vec{b}_8 =(0.0654, -0.6247, 0.7781),\;  \boxed{\vec{b}_9 = (-0.4556, -0.1679, 0.8742),\; \vec{b}_{10} =(0.4805, -0.0700, 0.8742)}$ & \\
 \hline\hline
\end{tabular}\label{tab:optimal1}
\end{table}
\end{widetext}
\twocolumngrid		

\begin{figure}[t]
		\centering
		\includegraphics[width=80mm]{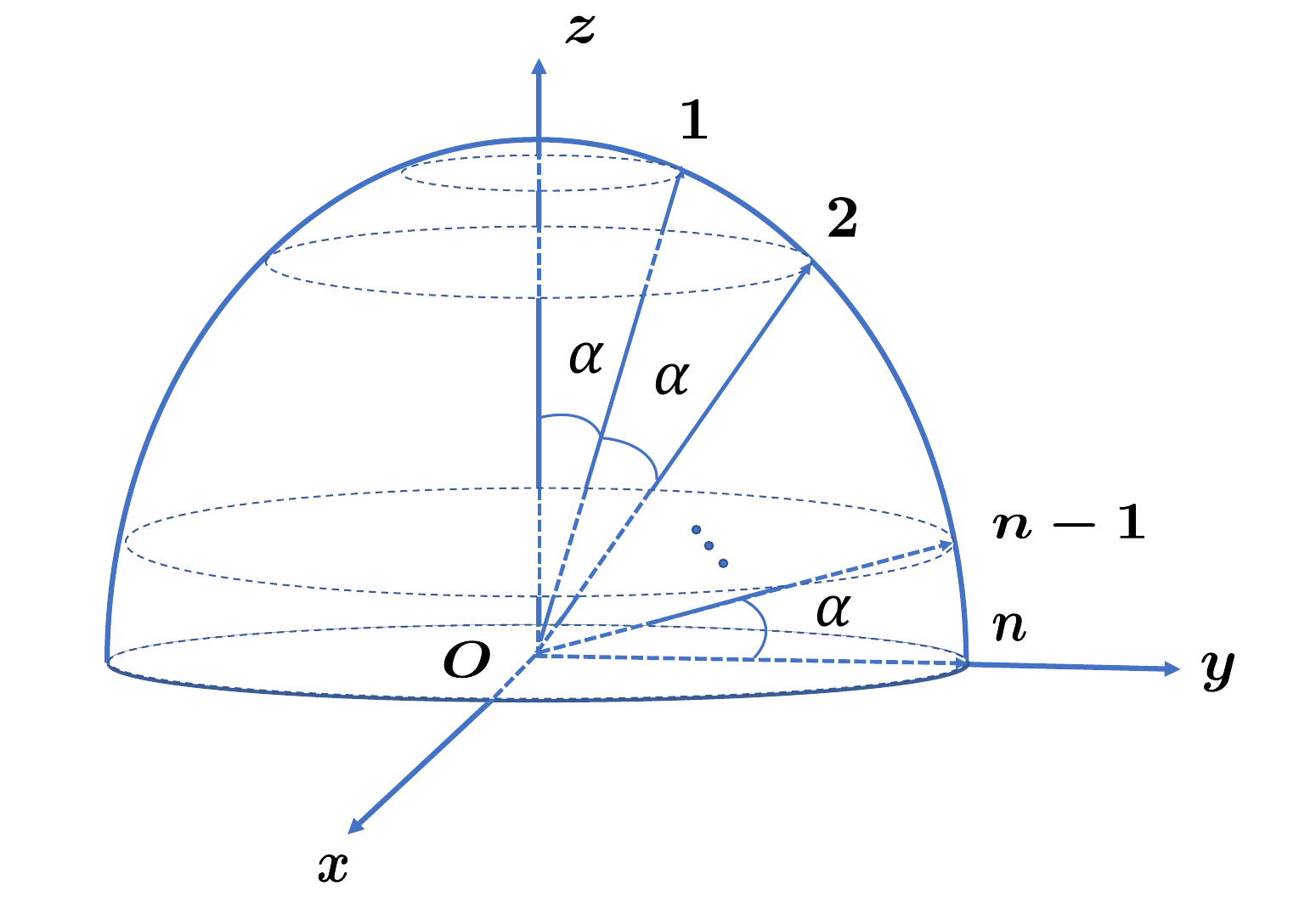}
		\caption{Illustration of the measurement directions $\vec{b}_j$  in northern hemisphere.}\label{fig:hemisphere}
	\end{figure}

We now come to study the optimal inequality for $N\rightarrow \infty$. In Fig. \ref{fig:hemisphere}, we have plotted the \emph{Northern Hemisphere}. Geometrically, given a Cartesian coordinate in a unit hemisphere with the origin $O$, then $\angle zOy=\pi/2$. First, we draw $(n-1)$ rays in the $zOy$ quadrant, such that $\angle zOy$ is uniformly divided into $n$ parts, i.e., $	\alpha={\pi}/(2n)$.
%		\begin{equation}
%			\alpha=\dfrac{\pi}{2n}.
%		\end{equation}
Second, based on which we portray $(n-1)$ circles $\big($denoted as $k=1,2,...,n-1$$\big)$ paralleling to the $xOy$ plane (i.e., the equatorial plane). Third, for the $k$-th circle,  we select $p_k$ points distributed evenly on its circumference, which correspond to $p_k$ vectors pointing from the origin $O$ to these points. For convenient, we mark these vectors as $\vec{r}_k^{\, j}$, ($j=1,2,...,p_k$). Soon one obtains the angle between the vector $\vec{r}_k^{\,j}$ and the $z$-axis as $k \alpha$,
%		\begin{equation}
%			k\,\theta=k\dfrac{\pi}{2\,n},
%		\end{equation}
and for the $k$-th circle, the sum of all vectors $\vec{r}_k^{\, j}$ reads
		\begin{equation}
			\sum_{j=1}^{p_k} \vec{r}_k^{\,j} =\vec{e}_z\, p_k \cos(k\alpha),
		\end{equation}
where $\vec{e}_z$ represents the unit vector along the $z$-axis.

Now, we regard $\vec{r}_k^{\, j}$ as one of the measurement directions of Bob; then totally we have the number of measurement settings as
$N=\sum_{k=1}^{n-1} p_k$,
%		\begin{equation}
%		N=\sum_{k=1}^{n-1} p_k,
%		\end{equation}
and the corresponding $N$-setting steering inequality is given in (\ref{eq:ineq}).
One can find that the sum of $N$ vectors on the whole hemisphere is
		\begin{equation}
			\vec{S}\equiv\sum_{k=1}^{n-1} \sum_{j=1}^{p_k} \vec{r}_k^{\, j}
			=\sum_{k=1}^{n-1} \vec{e}_z\, p_k\,\cos(k\alpha).
		\end{equation}
When $n\rightarrow \infty$ (i.e., $N\rightarrow \infty$), there are infinite points (or infinite measurement directions) of Bob. We choose them to evenly distribute on the northern hemisphere, then the ratio of $p_k/p_{k+1}$ is equal to $\ell_{k}/\ell_{k+1}$, where $\ell_k=2\pi R_k$, and $R_k=\sin(k\alpha)$ is the radius of the $k$-th circle, namely,
\begin{equation}
		p_k=P \sin(k\alpha),
\end{equation}
with $P$ being a constant number. The classical bound is given by
\begin{eqnarray}
		\mathcal{C}_N=\mathop{\rm max}\limits_{\{A_k^j\}}\left|\frac{1}{N}\left[\sum_{k=1}^{n-1} \sum_{j=1}^{p_k} (A_k^j \vec{r}_k^{\,j})\right]\right|.
\end{eqnarray}
In Table \ref{tab:optimal1} we have designed that the maximum of $\mathcal{C}_N$ occurs under the circumstance of all $A_k^j$ equal to 1.  Thus in the limit of $N\rightarrow \infty$, we obtain the LHS bound as
\begin{eqnarray}
		\lim_{N\rightarrow \infty}\mathcal{C}_N&=&\lim_{n\rightarrow \infty} \frac{|\vec{S}|}{N}=\lim_{n\rightarrow \infty} \frac{\sum_{k=1}^{n-1}  \sin k \alpha \cos k \alpha}{\sum_{k=1}^{n-1}  \sin k \alpha}\nonumber\\
&=&\lim_{n\rightarrow \infty} \frac{\cos \frac{\alpha}{2} \cdot \cos \frac{(n-1) \alpha}{2}}{\sqrt{2}} =\frac{1}{2}.
\end{eqnarray}
Therefore, by studying quantum violation of the infinity-setting EPR-steering inequality, we successfully achieve the desired critical value $V_{\rm c}^{\rm S}=\mathcal{C}_\infty=1/2$ for the two-qubit Werner state.

\emph{Conclusion.}---Quantum mechanics is essentially different from classical physics. This is manifested by the fact that the prediction of quantum correlations cannot be fully explained by classical models. The boundary between quantum mechanics and classical physics is a fundamental topic in quantum foundations, and the Werner state is a benchmark to test such a boundary. People have known the exact boundary between quantum mechanics and LHS model from the perspectives of projective measurements as well as positive operator-valued
measures~\cite{WJD07,WJD07PRA,2023ArxivZhang,2023ArxivRenner}, whereas how to demonstrate the critical value $V_{\rm c}^{\rm S}=1/2$ by the approach of steering inequality is a long-standing problem. In this work, by studying the optimal $N$-setting linear EPR-steering inequalities, we have successfully obtained the desired value $V_{\rm c}^{\rm S}=1/2$ for the two-qubit Werner state, thus resolving the long-standing problem. In SM \cite{SM}, we have also compared SJWP's inequalities and the optimal inequalities by using the generalized Werner states, the maximal entangled mixed states \cite{2003PRAWei}, and the all-versus-nothing states ~\cite{AVN2013}. One may finds that the latter has a stronger ability to detect the nonlocal property of quantum states than the former. Moreover, it is worth mentioning that the problem of the boundary between steering and LHS model may have connections with the problem of the boundary between compatible and incompatible measurements for depolarizing channels \cite{2022PRXQKu,2022NCKu}. Eventually, how to reveal the critical value $V_{\rm c}^{\rm B} =1/K_G(3)$ by Bell' inequality is still open, we shall investigate the problem in the near future.

\begin{acknowledgments}
   J.L.C. is supported by the National Natural Science Foundation of China (Grants No. 12275136 and 12075001) and the 111 Project of B23045. R.C.W, Z.C.L., X.R.X., and H.H.W. are supported by the Pilot Scheme of Talent Training in Basic Sciences (Boling Class of Physics, Nankai University), Ministry of Education. X.Y.F. is supported by the Nankai Zhide Foundation.		
\end{acknowledgments}

R.C.W, Z.C.L. and X.Y.F. contributed equally to this work.

	\end{document}

\renewcommand{\figureautorefname}{FIG.}
\renewcommand{\figurename}{FIG.}
\title{Supplemental Material for\\
``Revealing the Boundary between Quantum Mechanics and Classical Model by EPR-Steering Inequality''}

\author{Ruo-Chen Wang}
\affiliation{School of Physics, Nankai University, Tianjin 300071, People's Republic of China}

\author{Zhuo-Chen Li}
\affiliation{School of Physics, Nankai University, Tianjin 300071, People's Republic of China}

\author{Xing-Yan Fan}
\affiliation{Theoretical Physics Division, Chern Institute of Mathematics, Nankai
      University, Tianjin 300071, People's Republic of China}

\author{Xiang-Ru Xie}
\affiliation{School of Physics, Nankai University, Tianjin 300071, People's Republic of China}

\author{Hong-Hao Wei}
\affiliation{School of Physics, Nankai University, Tianjin 300071, People's Republic of China}

\author{Choo Hiap Oh}
\email{phyohch@nus.edu.sg}
\affiliation{Centre for Quantum Technologies and Department of Physics, National University of Singapore, 117543, Singapore}

\author{Jing-Ling Chen}
\email{chenjl@nankai.edu.cn}
\affiliation{Theoretical Physics Division, Chern Institute of Mathematics, Nankai
      University, Tianjin 300071, People's Republic of China}

\date{\today}
\maketitle

\tableofcontents

%\newpage

\section{Measurement Settings and LHS Bounds for Optimal Inequalities}

Here we list measurement settings and LHS bounds for the optimal EPR-steering inequalities from $N=2$ to $N=20$. For $N=2$ and $3$, the optimal inequalities are just SJWP's inequalities. For $N=3, 4, 5, 6$, we can obtain analytical expressions for measurement settings and LHS bounds, and for $N=7$ to $N=10$, we can obtain some quasi-analytical expressions for the measurement settings (see below).

\begin{itemize}
  \item \textcolor{blue}{\emph{The case of $N=2$}}. The $2$-setting SJWP inequality has been an optimal inequality. In Table I of main text, the measurement settings are given by
      \begin{eqnarray}
		&& \vec{b}_1 = (1, 0, 0), \;\;\;\; \vec{b}_2 = (0, 1, 0).
	\end{eqnarray}
However, in the Table II of main text, we have rotated them into the \emph{northern hemisphere}, which turn to
  \begin{eqnarray}\label{eq:b2}
		&& \vec{b}_1 = \left(0, -\frac{1}{\sqrt{2}}, \frac{1}{\sqrt{2}}\right), \;\;\;\; \vec{b}_2 = \left(0, \frac{1}{\sqrt{2}}, \frac{1}{\sqrt{2}}\right).
	\end{eqnarray}
Note that in Eq. (\ref{eq:b2}) we have designed these two vectors $\vec{b}_1$ and $\vec{b}_2$ in an elegant way, which have two nice properties: (i) The maximal value of $\mathcal{I}_2$ may occur at all $A_j$'s equal to $+1$, which means that the LHS bound can be calculated through $\mathcal{C}_2=|\vec{b}_1+\vec{b}_2|/2=1/\sqrt{2}\approx 0.7071$; (ii) Let $\vec{u}$ be the summation of all $\vec{b}_j$'s divided by $N=2$, then $\vec{u}$ points to the north pole, whose magnitude is just equal to $\mathcal{C}_2$, i.e., $\vec{u}=(\vec{b}_1+\vec{b}_2)/2=\hat{z}\, \mathcal{C}_2$.

  \item \textcolor{blue}{\emph{The case of $N=3$}}. The $3$-setting SJWP inequality has been an optimal EPR-steering inequality. In Table I of main text, the measurement settings read
      \begin{eqnarray}
		&& \vec{b}_1 = (1, 0, 0), \;\;\;\; \vec{b}_2 = (0, 1, 0), \;\;\;\; \vec{b}_3 = (0, 0, 1).
	\end{eqnarray}
Similarly, in the Table II of main text, we have rotated them into the \emph{northern hemisphere}, which turn to
    \begin{eqnarray}\label{eq:b3}
      && \vec{b}_1 = (\frac{\sqrt{6}}{3}, 0, \frac{\sqrt{3}}{3}),\qquad
      \vec{b}_2 = (-\frac{\sqrt{6}}{6}, \frac{\sqrt{2}}{2}, \frac{\sqrt{3}}{3}),\qquad
      \vec{b}_3 = (-\frac{\sqrt{6}}{6}, -\frac{\sqrt{2}}{2}, \frac{\sqrt{3}}{3}).
    \end{eqnarray}
    The maximal value of $\mathcal{I}_3$ may occur at all $A_j$'s equal to $+1$,  and the LHS bound can be calculated through
    $\vec{u}=(\vec{b}_1+\vec{b}_2+\vec{b}_3)/3=\hat{z}\, \mathcal{C}_3$, with $\mathcal{C}_3=1/\sqrt{3}\approx 0.5773$.

  \item \textcolor{blue}{\emph{The case of $N=4$}}. In the Table II of main text, the measurement settings for optimal inequality are given by
    \begin{eqnarray}\label{eq:b4}
      && \vec{b}_1 = (\frac{2}{\sqrt{5}},0, \frac{1}{\sqrt{5}}),\;\;\;
      \vec{b}_2 = (-\frac{1}{2\sqrt{5}}, \frac{\sqrt{3}}{2}, \frac{1}{\sqrt{5}}),\;\;\;
      \vec{b}_3 = (-\frac{7}{4\sqrt{5}}, -\frac{\sqrt{3}}{4}, \frac{1}{\sqrt{5}}),\;\;\;
      \vec{b}_4 =  (\frac{1}{4\sqrt{5}}, -\frac{\sqrt{3}}{4}, \frac{2}{\sqrt{5}}).
    \end{eqnarray}
    The maximal value of $\mathcal{I}_4$ may occur at all $A_j$'s equal to $+1$,  and the LHS bound can be calculated through
    $\vec{u}=(\vec{b}_1+\vec{b}_2+\vec{b}_3+\vec{b}_4)/4=\hat{z}\, \mathcal{C}_4$, with $\mathcal{C}_4=\sqrt{5}/4\approx 0.5590$.

  \item \textcolor{blue}{\emph{The case of $N=5$}}. The measurement settings (not in the \emph{northern hemisphere}) for optimal inequality read
  \begin{eqnarray}\label{eq:b5a}
      && \vec{b}_1 = (0,0,1),\qquad\vec{b}_2 = (1,0,0),\qquad
      \vec{b}_3 =\Bigg(\frac{1}{4}(-3+\sqrt{33}), \sqrt{-\frac{13}{8}+\frac{3 \sqrt{33}}{8}}, 0\Bigg), \notag \\
      && \vec{b}_4 = \Bigg(\frac{1}{8}(3-\sqrt{33}),
          \frac{1}{4} \sqrt{\frac{3}{2}(1+\sqrt{33})},-\frac{1}{2}\Bigg),\qquad  \vec{b}_5 =\Bigg(\frac{1}{8}(3-\sqrt{33}),
          \frac{1}{4} \sqrt{\frac{3}{2}(1+\sqrt{33})},\frac{1}{2}\Bigg).
    \end{eqnarray}
  One may rotate them to the northern hemisphere as
    \begin{eqnarray}\label{eq:b5b}
      && \vec{b}_1 = (-0.9297 , 0 , 0.3683),\qquad\vec{b}_2 = (0.1459 , -0.9182 , 0.3683),\qquad  \vec{b}_5 = (0.6837 , 0.6300 , 0.3683), \notag \\
      && \vec{b}_4 = (-0.2460 , 0.6300 , 0.7366),  \qquad \vec{b}_3 = (0.3461 , -0.3418 , 0.8737).
    \end{eqnarray}
    (Note that the analytical expression in the northern hemisphere is lengthy, thus we have written them in a form as in Eq. (\ref{eq:b5b})).
     The maximal value of $\mathcal{I}_5$ may occur at all $A_j$'s equal to $+1$,  and the LHS bound can be calculated through
    $\vec{u}=(\vec{b}_1+\vec{b}_2+\vec{b}_3+\vec{b}_4+\vec{b}_5)/5=\hat{z}\, \mathcal{C}_5$, with $\mathcal{C}_5= \sqrt{(9+\sqrt{33})/50} \approx 0.5430$.

  \item \textcolor{blue}{\emph{The case of $N=6$}}. The measurement settings for optimal inequality are given by
    \begin{eqnarray}\label{eq:b6}
      && \vec{b}_1 = \bigg(\frac{3}{\sqrt{10}}, 0, \frac{1}{\sqrt{10}}\bigg),\qquad
        \vec{b}_2 = \bigg(-\frac{1}{3\sqrt{10}},\frac{2\sqrt{2}}{3},\frac{1}{\sqrt{10}}\bigg),\qquad
        \vec{b}_3 =\bigg(-\sqrt{\frac{2}{5}}, -\frac{1}{\sqrt{2}}, \frac{1}{\sqrt{10}}\bigg), \notag \\
      && \vec{b}_4 =\bigg(-\frac{7}{3\sqrt{10}},\frac{1}{3\sqrt{2}},\frac{2}{\sqrt{10}}\bigg),\qquad
        \vec{b}_5 = \bigg(\frac{1}{\sqrt{10}}, -\frac{1}{\sqrt{2}}, \frac{2}{\sqrt{10}}\bigg),\qquad
        \vec{b}_6 = \bigg(\frac{1}{3} \sqrt{\frac{2}{5}}, \frac{1}{3\sqrt{2}}, \frac{3}{\sqrt{10}}\bigg).
    \end{eqnarray}
    The maximal value of $\mathcal{I}_6$ may occur at all $A_j$'s equal to $+1$,  and the LHS bound can be calculated through
    $\vec{u}=(\vec{b}_1+\vec{b}_2+\vec{b}_3+\vec{b}_4+\vec{b}_5+\vec{b}_6)/6=\hat{z}\, \mathcal{C}_6$, with $\mathcal{C}_6=\sqrt{10}/6 \approx 0.5270$.

  \item \textcolor{blue}{\emph{The case of $N=7$}}. The measurement settings for optimal inequality (not in the \emph{northern hemisphere}) can be the following form
    \begin{eqnarray}\label{eq:b7}
      && \vec{b}_1 = (1, 0, 0) ,\qquad
        \vec{b}_2 = (\sin\theta_2 \cos\phi_2, \sin\theta_2 \sin\phi_2, \cos\theta_2),\qquad
        \vec{b}_3 = \left(\frac{1}{\sqrt{2}}, 0, \frac{1}{\sqrt{2}}\right), \notag \\
      && \vec{b}_4 =(\cos\phi_4, \sin\phi_4, 0),\qquad
        \vec{b}_5 = (\sin\theta_2 \cos\phi_2, \sin\theta_2 \sin\phi_2, -\cos\theta_2),\qquad
        \vec{b}_6 =  (0, 0, 1), \nonumber\\
        && \vec{b}_7= \left(\frac{1}{\sqrt{2}}, 0, \frac{1}{\sqrt{2}}\right),
    \end{eqnarray}
    where
    \begin{eqnarray}\label{eq:b7-2}
      && \phi_4 \approx 5.249546515851768, \;\;\; \phi_2 \approx -1.8564089024476422, \;\;\; \theta_2 \approx -4.280275193917692,
      \end{eqnarray}
    and they satisfy following relations
    \begin{eqnarray}\label{eq:b7-1}
      && \cos\phi_4+2 \cos\phi_2\sin\theta_2=0, \nonumber\\
      && \left[\left(1+\sqrt{2}\right) \cos\phi_2+\cos(\phi_2-\phi_4)\right] \sin\theta_2=0, \nonumber\\
      && 2 \left(1+\sqrt{2}\right) \cos\theta_2-2 (\cos\theta_2)^2 +\cos\phi_4 \left(\sqrt{2}+2 \cos\phi_2\sin\theta_2\right)\nonumber\\
      && \;\;\;\;+2 \sin\theta_2 \left[(1+\sqrt{2})\cos\phi_2+ (1+\sin\phi_4)\sin\theta_2 \right]=0.
      \end{eqnarray}
    By the way, from the first two equations of (\ref{eq:b7-1}) one can have
     \begin{eqnarray}
      && \phi_2=-\arccos\left[\frac{\sin\phi_4}{\sqrt{4+2 \sqrt{2}+2 (1+\sqrt{2}) \cos\phi_4}}\right], \nonumber\\
      && \theta_2=-\pi + \arcsin\left[\frac{\cos\phi_4}{2\cos\phi_2}\right],
      \end{eqnarray}
    and $\phi_4$ takes $5.249546515851768$ as shown in Eq. (\ref{eq:b7-2}), which is very closed to the analytical value $2\pi-\arccos\left[\frac{27-\sqrt{2}}{50}\right]\approx5.24957$.

    One may rotate these seven vectors to the northern hemisphere, then they become
    \begin{eqnarray}\label{eq:b7}
      && \vec{b}_1 = (0.811031, 0.518332, 0.271220) ,\qquad
        \vec{b}_2 = (-0.962517, 0, 0.271220),\qquad \nonumber\\
     &&   \vec{b}_3 = (-0.025624, -0.942341, 0.333670), \qquad
      \vec{b}_4 =(0.653592, -0.508171, 0.560874),\qquad \nonumber\\
      &&  \vec{b}_5 =(-0.550186, 0.518331, 0.654697),\qquad
        \vec{b}_6 = (0.184470, 0.733036, 0.654697), \nonumber\\
        && \vec{b}_7=  (-0.110765, -0.319187, 0.941196).
    \end{eqnarray}
    (Here we have renumbered the vectors, such that $\vec{b}_1$ is closed to the equatorial plane and $\vec{b}_7$ is far from the equatorial plane). The maximal value of $\mathcal{I}_7$ may occur at all $A_j$'s equal to $+1$,  and the LHS bound can be calculated through
    $\vec{u}=(\vec{b}_1+\vec{b}_2+\vec{b}_3+\vec{b}_4+\vec{b}_5+\vec{b}_6+\vec{b}_7)/7=\hat{z}\, \mathcal{C}_7$, with
     \begin{eqnarray}
      && \mathcal{C}_7=\frac{1}{7} \sqrt{4+2\sqrt{2}+(2\sin\theta_2\sin\phi_2+\sin\phi_4)^2}\approx 0.562784.
      \end{eqnarray}

  \item \textcolor{blue}{\emph{The case of $N=8$}}. The measurement settings for optimal inequality (not in the \emph{northern hemisphere}) can be the following form
      \begin{eqnarray}
      && \vec{b}_1 = \left(-\frac{1}{\sqrt{2}}, 0, \frac{1}{\sqrt{2}}\right),\qquad
        \vec{b}_2 = (\cos\phi_2, \sin\phi_2, 0),\qquad\vec{b}_3 = ( 0,0,1), \notag \\
      && \vec{b}_4 = (-1,0,0),\qquad\vec{b}_5 = (\cos\phi_5, \sin\phi_5, 0),\qquad
        \vec{b}_6 = \left(\frac{4}{5}   \cos\phi_6, \frac{4}{5}   \sin\phi_6,-\frac{3}{5}\right), \notag \\
      && \vec{b}_7 = \left(\frac{4}{5}   \cos\phi_6, \frac{4}{5}   \sin\phi_6, \frac{3}{5}\right), \qquad\vec{b}_8 = \left( \frac{1}{\sqrt{2}}, 0, \frac{1}{\sqrt{2}}\right),
    \end{eqnarray}
    where
    \begin{eqnarray}
      && \phi_2 \approx 2.2542981930847485,\qquad\phi_5 \approx1.0982166105825837,\qquad\phi_6 \approx 1.4603674317905544,
    \end{eqnarray}
    and they satisfy following relations
    \begin{eqnarray}
      && \cos\phi_2 +\cos\phi_5 + \frac{8}{5} \cos\phi_6=0,\\
      && (1+\sqrt{2})\cos\phi_2 +\cos(\phi_2 -\phi_5) +\frac{8}{5} \cos(\phi_2 -\phi_6)=0, \\
      && (\cos\phi_2 -\cos\phi_5)+(\sin\phi_2 +\sin\phi_5)^2 +\frac{8}{5} (\sin\phi_2 +\sin\phi_5)\sin\phi_6 =\frac{66+30\sqrt{2}}{25}.
    \end{eqnarray}
     One may rotate these eight vectors to the northern hemisphere, then they become
     \begin{eqnarray}
      && \vec{b}_1 = (0.968666, 0.065740, 0.239509),\qquad\;\;
        \vec{b}_2 = (-0.747121, -0.620037, 0.239509),\notag \\
      && \vec{b}_3 = (-0.209859, 0.942222, 0.261106),\qquad
        \vec{b}_4 = (-0.650221, 0.492982, 0.578084), \notag \\
      && \vec{b}_5 = (0.622779, -0.527067, 0.578227),\qquad
        \vec{b}_6 = (-0.087923, -0.811125, 0.578227), \notag \\
      && \vec{b}_7 = (0.431679, 0.492111, 0.755963),\qquad\;\;
        \vec{b}_8= (-0.328001, -0.034827, 0.944574).
    \end{eqnarray}
    (Here we have renumbered the vectors, such that $\vec{b}_1$ is closed to the equatorial plane and $\vec{b}_8$ is far from the equatorial plane). The maximal value of $\mathcal{I}_8$ may occur at all $A_j$'s equal to $+1$,  and the LHS bound can be calculated through
    $\vec{u}=(\vec{b}_1+\vec{b}_2+\vec{b}_3+\vec{b}_4+\vec{b}_5+\vec{b}_6+\vec{b}_7+\vec{b}_8)/8=\hat{z}\, \mathcal{C}_8$, with
    with the LHS bound obtained  through $\vec{u}=(\vec{b}_1+\vec{b}_2+\vec{b}_3+\vec{b}_4+\vec{b}_5+\vec{b}_6+\vec{b}_7 +\vec{b}_8)/8=\hat{z}\, \mathcal{C}_8$, and
    \begin{eqnarray}
      && \mathcal{C}_8 =\sqrt{\frac{1}{800} \Bigl\{107 + 25 \sqrt{2} + 25 \cos(\phi_2 - \phi_5)+40\big[\cos(\phi_2 - \phi_6)+\cos(\phi_5 - \phi_6)\big]\Bigr\}}\approx 0.521867.
    \end{eqnarray}

    \item \textcolor{blue}{\emph{The case of $N=9$}}. The measurement settings for optimal inequality (not in the \emph{northern hemisphere}) can be the following form
         \begin{eqnarray}
      && \vec{b}_1 =(\sin\theta_1, 0,\cos\theta_1) ,\qquad \vec{b}_2 =(\sin\theta_2\,\cos\phi_2, \sin\theta_2\sin\phi_2,\cos\theta_2),\qquad\vec{b}_3 =(0, 0, 1), \notag \\
      && \vec{b}_4 =(-\sin\theta_1\,\cos 2 \phi_7,\:-\sin\theta_1\,\sin2\phi_7,\:\cos\theta_1),\qquad
        \vec{b}_5 =(-\sin\theta_1\,\cos( 2 \phi_7-\phi_8),-\sin\theta_1\,\sin( 2 \phi_7-\phi_8),\cos\theta_1),\notag \\
      && \vec{b}_6 =(-\sin\theta_2\,\cos( 2 \phi_7-\phi_2),-\sin\theta_2\,\sin( 2 \phi_7-\phi_2),\cos\theta_2),\qquad
        \vec{b}_7 =(\sin\theta_7\,\cos\phi_7, \sin\theta_7\sin\phi_7,\cos\theta_7), \notag \\
      && \vec{b}_8 =(\sin\theta_1\,\cos\phi_8, \sin\theta_1\sin\phi_8,\cos\theta_1),\qquad\vec{b}_9 = (-\sin\theta_7\,\cos\phi_7, -\sin\theta_7\sin\phi_7,\cos\theta_7),
    \end{eqnarray}
    where
    \begin{eqnarray}
      && \theta_1 \approx 1.1511827283784763,\qquad
        \theta_2 \approx 1.3594601487486428,\qquad
        \theta_7 \approx 0.6221259831692709, \notag \\
      && \phi_2 \approx -4.205370916319264, \qquad
       \phi_7 \approx -1.1680009532610969, \qquad
       \phi_8 \approx -3.158241159860658.
    \end{eqnarray}
     One may rotate these ten vectors to the northern hemisphere, then they become
    \begin{eqnarray}
      && \vec{b}_1 = (-0.946974, -0.239896, 0.213751),\qquad
        \vec{b}_2 =(0.975285, 0.055948, 0.213751), \notag \\
      && \vec{b}_3 =(0.223587, 0.900606, 0.372717),\qquad
        \vec{b}_4= (-0.484042, 0.791698, 0.372717), \notag \\
      && \vec{b}_5 =(0.49183, -0.750648, 0.441170) ,\qquad
        \vec{b}_6 =(-0.243361, -0.863797, 0.441170), \notag \\
      && \vec{b}_7 =(0.570929, 0.121485, 0.811962),\qquad
        \vec{b}_8 =(-0.581037, -0.055808, 0.811961), \notag \\
      && \vec{b}_9 = (-0.006217, 0.040411, 0.999164).
    \end{eqnarray}
    (Here we have renumbered the vectors, such that $\vec{b}_1$ is closed to the equatorial plane and $\vec{b}_{9}$ is far from the equatorial plane).
    The maximal value of $\mathcal{I}_9$ may occur at all $A_j$'s equal to $+1$,  and the LHS bound can be calculated through
    $\vec{u}=(\vec{b}_1+\vec{b}_2+\vec{b}_3+\vec{b}_4+\vec{b}_5+\vec{b}_6+\vec{b}_7+\vec{b}_8+\vec{b}_9)/9=\hat{z}\, \mathcal{C}_{9}$, with
    \begin{eqnarray}
      && \mathcal{C}_9 \approx 0.519818.
    \end{eqnarray}

    \item \textcolor{blue}{\emph{The case of $N=10$}}. The measurement settings for optimal inequality (not in the \emph{northern hemisphere}) can be the following form
         \begin{eqnarray}
      && \vec{b}_1 = (\sin\theta_1\,\cos\phi_1,\:\sin\theta_1\,\sin\phi_1,\:\cos\theta_1),\qquad\vec{b}_2 = (1,0,0),\qquad\vec{b}_3 =(\cos\phi_3,\sin\phi_3,0) \notag \\
      && \vec{b}_4 =\big(\sin\theta_1\,\cos(2\,\phi_3 -\phi_1),\:
          \sin\theta_1\,\sin(2\,\phi_3 -\phi_1),\:\cos\theta_1\big),\qquad
        \vec{b}_5 =(0,0,1),\notag \\
      && \vec{b}_6 =\big(\cos(2\,\phi_3),\:\sin(2\,\phi_3),\:0\big),\qquad
        \vec{b}_7 =(\cos\phi_7,\sin\phi_7,0), \notag \\
      && \vec{b}_8 =(-\sin\theta_1\,\cos\phi_1,\:-\sin\theta_1\,\sin\phi_1,\:\cos\theta_1),\qquad\vec{b}_9 =\big(\cos(2\,\phi_3 -\phi_7),\:\sin(2\,\phi_3 -\phi_7),\:0\big), \notag \\
      && \vec{b}_{10} =\big(-\sin\theta_1\,\cos(2\,\phi_3 -\phi_1),\:-\sin\theta_1\,\sin(2\,\phi_3 -\phi_1),\:\cos\theta_1\big),
    \end{eqnarray}
    where
    \begin{eqnarray}
      && \theta_1 \approx 0.7146287054686191,\qquad
        \phi_1 \approx 2.0452288433223678,\qquad
        \phi_3 \approx 1.244110381691307, \notag \\
      && \phi_7 \approx 1.8839410726395784.
    \end{eqnarray}
     One may rotate these ten vectors to the northern hemisphere, then they become
    \begin{eqnarray}
      && \vec{b}_1 = (0.915980, 0.346916, 0.201568),\qquad
        \vec{b}_2 =(-0.967957, 0.149768, 0.201568), \notag \\
      && \vec{b}_3 =(-0.381747, -0.873784, 0.301280),\qquad
        \vec{b}_4= (0.554376, -0.775821, 0.301280), \notag \\
      && \vec{b}_5 = (0.528849, 0.682951, 0.503881),\qquad
        \vec{b}_6 =(-0.658780, 0.558670, 0.503881), \notag \\
      && \vec{b}_7 =(-0.080984, 0.773887, 0.628125),\qquad
        \vec{b}_8 =(0.065374, -0.624714, 0.778112), \notag \\
      && \vec{b}_9 = (-0.455624, -0.167912, 0.874193),\qquad
        \vec{b}_{10} =(0.480513, -0.069961, 0.874193).
    \end{eqnarray}
    %(Here we have renumbered the vectors, such that $\vec{b}_1$ is closed to the equatorial plane and $\vec{b}_{10}$ is far from the equatorial plane).
    The maximal value of $\mathcal{I}_{10}$ may occur at all $A_j$'s equal to $+1$,  and the LHS bound can be calculated through
    $\vec{u}=(\vec{b}_1+\vec{b}_2+\vec{b}_3+\vec{b}_4+\vec{b}_5+\vec{b}_6+\vec{b}_7+\vec{b}_8+\vec{b}_9+\vec{b}_{10})/10=\hat{z}\, \mathcal{C}_{10}$, with
    \begin{eqnarray}
      && \mathcal{C}_{10} = \frac{1}{\sqrt{50}} \times\nonumber\\
      && \sqrt{7 + 4\cos\theta_1 + 4\cos2\theta_1 + 2\cos\phi_3 +\cos2\phi_3+ 2\cos(\phi_3 - \phi_7) +\cos[2 (\phi_3 - \phi_7)]
      + 2\cos(2\phi_3 - \phi_7) +2\cos\phi_7} \nonumber\\
      && \approx 0.516808.
    \end{eqnarray}
\end{itemize}

 For $N=12$ to $N=20$, the numerical results are listed in the following Table \ref{tab:optimal2}.

\begin{table}[ht]
	\centering
\caption{Measurement settings and the corresponding LHS bounds for optimal inequalities. Here $N=12, 14, 16, 18, 20$.}
\begin{tabular}{llc}
\hline\hline
$N$\;\;\;\; &  Measurement directions $\vec{b}_j$   & \;\;LHS bound $\mathcal{C}_N$   \\
  \hline
 12 & $\vec{b}_1 = (-0.8002, -0.5577, 0.22050),\; \vec{b}_2 =(-0.859, 0.4357,
  0.2686),$ & $0.5124$ \\
    & $\boxed{\vec{b}_3 = (-0.2008, 0.9394, 0.2779),\; \vec{b}_4 = (0.9529, 0.1212, 0.2779), \vec{b}_{5} =(0.7407, -0.6117, 0.2779)},$ &  \\
    & $\vec{b}_6=(-0.1576, -0.9277, 0.3383),\; \vec{b}_7 =(0.5093, 0.7663, 0.3917),\; \vec{b}_{8} = (0.2982, -0.5917, 0.7490),$ &  \\
    & $\vec{b}_9 =(-0.6580, 0.0164, 0.7529) , \; \vec{b}_{10} =(-0.1256, 0.5548, 0.8225),\;   $ &  \\
    & $ \;\vec{b}_{11} = (0.5214, 0.1453, 0.8409),\;\vec{b}_{12} =(-0.2212, -0.2903, 0.9310)$ & \\
     \hline
 14 & $\vec{b}_1 = (-0.4540 ,  -0.8654  ,  0.2122),\; \vec{b}_2 = (0.0134 ,   0.3709  ,  0.9286),\;  \vec{b}_3 = (0.2711 , -0.9230 ,   0.2732) , $ & $0.5103$ \\
    & $\vec{b}_4 =(  -0.8330  , -0.2467 ,   0.4953), \; \vec{b}_5 =(0.9506  , 0.2742   ,0.1449) ,\; \vec{b}_6 =( -0.9056  ,  0.4002 ,  0.1400) ,$ &  \\
    & $\vec{b}_7 = ( -0.0360  ,  0.9049,    0.4240), \; \vec{b}_8 =( 0.5655,    0.1908,    0.8024) , \; \vec{b}_9 =(-0.1636  , -0.6802  ,  0.7145) $ &  \\
    & $\vec{b}_{10} =(0.7675  , -0.4892 ,   0.4142), \; \vec{b}_{11} = (0.6443 ,   0.6676  ,  0.3729), \; \vec{b}_{12} =(-0.5659  , 0.6873  ,  0.4553) $ &  \\
   & $\vec{b}_{13} =(-0.5152  ,  0.0297  ,  0.8566), \; \vec{b}_{14} =( 0.2608  , -0.3214  ,  0.9103)  $ &  \\
      \hline
 16 & $\vec{b}_1 = (-0.4330 ,  -0.8834 ,   0.1790),\; \vec{b}_2 =( 0.5947   , 0.3580  ,  0.7198) ,\;  \vec{b}_3 = (-0.9199 ,   0.3544 ,   0.1679), $ & $0.5096$ \\
    & $\vec{b}_4 =(0.6250  , -0.6651 ,   0.4088), \; \vec{b}_5 =(-0.8531  , -0.2469 ,   0.4596) ,\; \vec{b}_6 =(0.0640 ,   0.6012  ,  0.7965) ,$ &  \\
    & $\vec{b}_7 = (0.3379 ,   0.8924   , 0.2991), \; \vec{b}_8 = (0.4152  , -0.2369   , 0.8784), \; \vec{b}_9 =(-0.6113 ,   0.4624   , 0.6422) $ &  \\
    & $\vec{b}_{10} =(-0.2007,  -0.6999  ,  0.6854), \; \vec{b}_{11} = (-0.4692  ,  0.8396 ,   0.2736), \; \vec{b}_{12} =(-0.0471,    0.0997  ,  0.9939) $ &  \\
   & $\vec{b}_{13} =(-0.4821 ,  -0.2496 ,   0.8398), \; \vec{b}_{14} =(0.8921  ,  0.4315 ,   0.1339) , \; \vec{b}_{15} =(0.1751 ,  -0.9430,    0.2830) $ &  \\
     & $\vec{b}_{16} =( 0.9124  , -0.1144 ,   0.3931) $ &  \\
       \hline
 18 & $\vec{b}_1 = (-0.3998 ,   0.5864  ,  0.7044),\; \vec{b}_2 =(-0.1977 ,  -0.2985 ,   0.9337) ,\;  \vec{b}_3 = (-0.3298 ,   0.2137 ,   0.9195), $ & $0.5082$ \\
    & $\vec{b}_4 =(0.6925 ,  -0.4245  ,  0.5833), \; \vec{b}_5 =(0.4484 ,   0.8649 ,   0.2256) ,\; \vec{b}_6 =( 0.1220 ,   0.7780,    0.6163) ,$ &  \\
    & $\vec{b}_7 =( 0.1244  , -0.6324,    0.7646) , \; \vec{b}_8 = ( -0.6623 ,  -0.4853 ,   0.5709), \; \vec{b}_9 =(0.9697 ,  -0.2072 ,   0.1294) $ &  \\
    & $\vec{b}_{10} =(-0.1528  , -0.9243  ,  0.3498), \; \vec{b}_{11} = ( 0.4199 ,  -0.8794  ,  0.2245), \; \vec{b}_{12} =(-0.8173 ,   0.4916 ,   0.3007) $ &  \\
   & $\vec{b}_{13} =( -0.7830 ,  -0.6046   , 0.1457), \; \vec{b}_{14} =(0.4634  ,  0.3948 ,   0.7933) , \; \vec{b}_{15} = (0.8772 ,   0.2919 ,   0.3811)$ &  \\
     & $\vec{b}_{16} =( -0.3433   , 0.9269  ,  0.1515), \; \vec{b}_{17} =(-0.8849 ,  -0.0570 ,   0.4623) , \; \vec{b}_{18} =( 0.4534 ,  -0.0348 ,   0.8907) $ &  \\
   \hline
 20 & $\vec{b}_1 = ( -0.4771  , -0.3132 ,   0.8212),\; \vec{b}_2 =(-0.9316  ,  0.2859  ,  0.2241) ,\;  \vec{b}_3 = (-0.8297  , -0.0335 ,   0.5572), $ & $0.5073$ \\
 & $\vec{b}_4 =(0.7434,  -0.0761   , 0.6645), \; \vec{b}_5 =(-0.0856 ,  -0.9817  ,  0.1703) ,\; \vec{b}_6 =(-0.4585  , -0.7249   , 0.5140) ,$ &  \\
 & $\vec{b}_7 =(  0.8713  , -0.4185  ,  0.2562) , \; \vec{b}_8 = (  -0.5864  ,  0.7777    ,0.2262), \; \vec{b}_9 =(0.4133 ,   0.3046 ,   0.8581) $ &  \\
 & $\vec{b}_{10} =(0.4841 ,  -0.8210,    0.3025), \; \vec{b}_{11} = ( 0.3804  , -0.4887   , 0.7852), \; \vec{b}_{12} =( -0.0581 ,   0.0345 ,   0.9977) $ &  \\
 & $\vec{b}_{13} =(0.0123 ,   0.9731,    0.2302), \; \vec{b}_{14} =(-0.8229 ,  -0.5475 ,  0.1518) , \; \vec{b}_{15} = (0.3358 ,   0.6873 ,   0.6441)$ &  \\
 & $\vec{b}_{16} =( 0.0713,   -0.6120,    0.7876), \; \vec{b}_{17} =(-0.3037  ,  0.6792   , 0.6681) , \; \vec{b}_{18} =(0.9445,    0.2249  ,  0.2395) $ &  \\
 & $\vec{b}_{19} =( -0.3920  ,  0.3529 ,   0.8496), \; \vec{b}_{20} =(0.6895  ,  0.6968  ,  0.1978)$ &  \\
 %  \hline
% & & \\
 \hline\hline
\end{tabular}\label{tab:optimal2}
\end{table}

	\section{Comparisons between SJWP's inequalities and Optimal Inequalities}

	\subsection{Generalized Werner State}

The generalized Werner state is given by
 \begin{eqnarray}
		&& \rho_{\rm GW} = V |{\Phi_-(\theta)}\rangle\langle{\Phi_-(\theta)}|  + (1-V)\frac{\mathbb{I}}{4},
	\end{eqnarray}
    where
    \begin{eqnarray}
		&& |{\Phi_-(\theta)}\rangle=\cos\theta |{01}\rangle -\sin\theta|{10}\rangle
	\end{eqnarray}
    is the non-maximally entangled state of two qubits, and $\theta\in[0,\pi/2],\;V\in[0,1]$. When $\theta=\pi/4$, $|{\Phi_-(\theta)}\rangle$ reduces to the maximally entangled state and accordingly the generalized Werner state reduces to the usual Werner state. In the matrix form, the state $\rho_{\rm GW}$ reads
    \begin{eqnarray}
\rho_{\rm GW}=\left[ \begin{array}{cccc}
              \frac{1-V}{4} & 0 & 0 & 0   \\
             0 & V\cos^{2}\theta +\frac{1-V}{4} & -V\sin \theta \cos \theta  & 0 \\
             0 & -V\sin \theta \cos \theta & V\sin^{2}\theta +\frac{1-V}{4} & 0 \\
             0  & 0 & 0 & +\frac{1-V}{4}
           \end{array}\right].
\label{colorN}
\end{eqnarray}

\begin{figure}[h]
		\centering
\includegraphics[width=80mm]{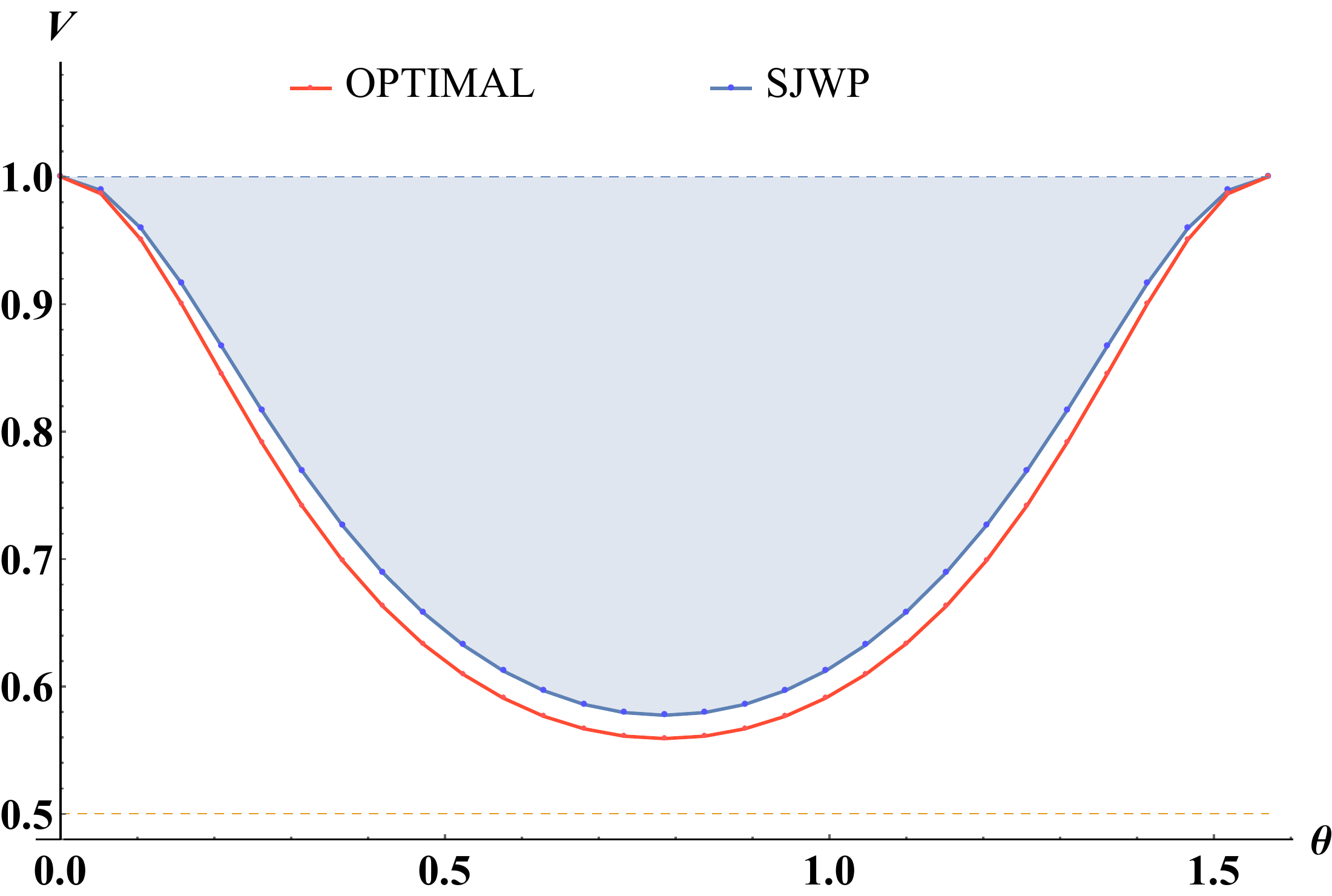}\hspace{3mm}
\includegraphics[width=80mm]{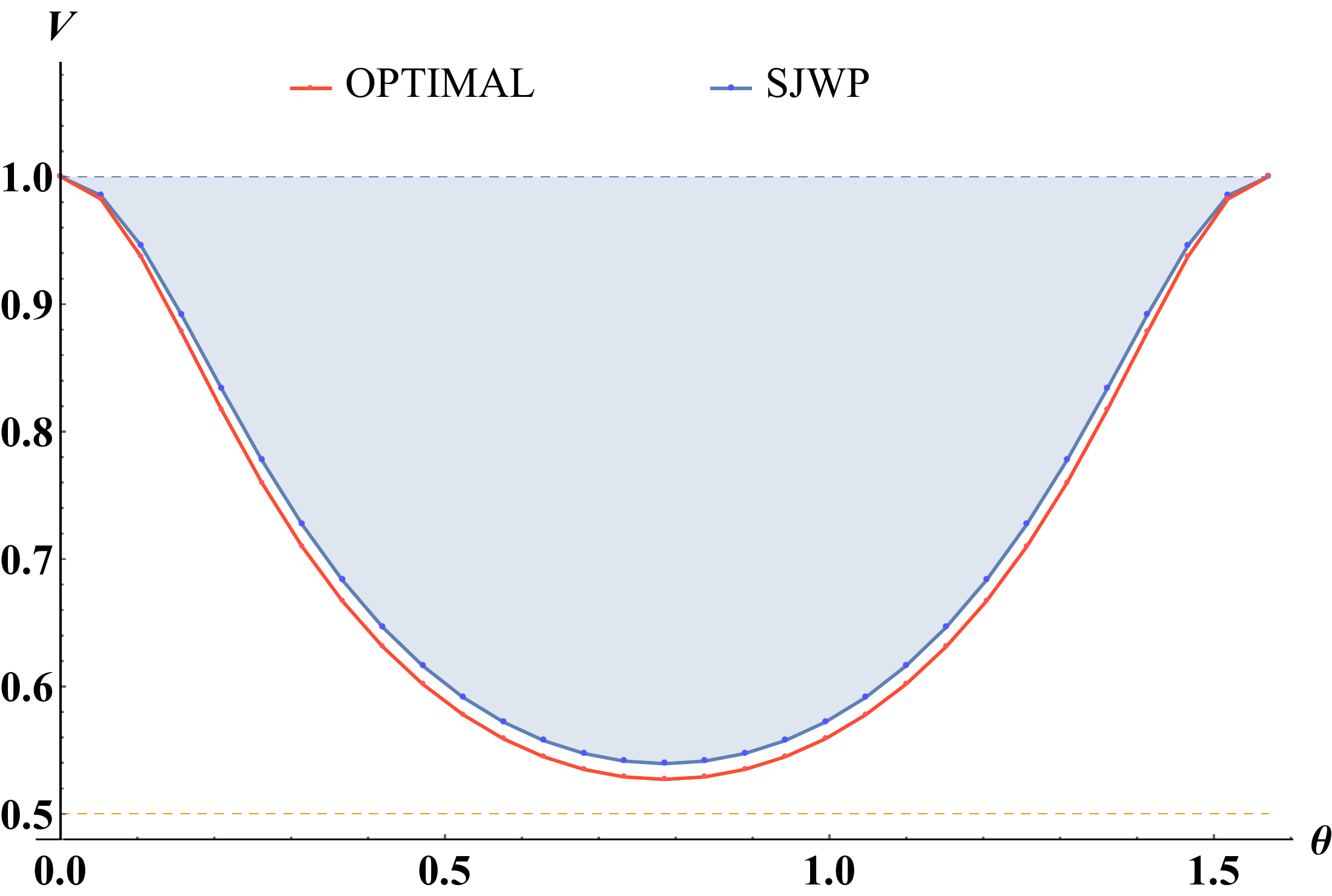}
		\caption{Quantum violations of SJWP's inequalities and optimal inequalities for generalized Werner state (left: 4-setting; right: 6-setting). The latter can detect more regions of $(V, \theta)$ than the former.}\label{fig:GWS}
\end{figure}

In Fig. \ref{fig:GWS}, we have compared quantum violations of SJWP's inequalities and optimal inequalities for generalized Werner state. We find that the detection ability of the latter is stronger than that of the former.

    \subsection{Maximal Entangled Mixed State}

    The maximal entangled mixed state (MEMS) reads \cite{2003PRAWei}
\begin{eqnarray}
\rho_{\rm MEMS}=\left[ \begin{array}{cccc}
             g & 0 & 0 & \frac{\gamma}{2}    \\
             0 & 1-2g & 0 & 0 \\
             0 & 0 & 0 & 0 \\
             \frac{\gamma}{2}  & 0 & 0 & g
           \end{array}\right],
\end{eqnarray}
with $g=1/3$ for $0\leq \gamma\leq 2/3$, and $g=\gamma/2$ for $2/3<
\gamma\leq 1$.

%		\begin{table}[h]
%			\centering
%			\caption{}
%			\begin{tabular}{ccccccccccccccc}
%				\hline\hline
%				$N$ & 2      & 3      & 4      & 5     & 6      & 7      & 8      & 9      & 10     & 12     & 14     & 16     & 18     & 20     \\\hline
%				SJWP & 0.7072 \,& 0.6667 \,& 0.6667 \,& /     & 0.6243 \,& /      & /      & /      & 0.6029 \,&
%				
%				/      & /      & /      & /      & /      \\
%				OPTIMAL \,& 0.7072 \, & 0.6667 \, & 0.6319 \, & 0.618 \, & 0.6047 \, & 0.6033 \, & 0.5972 \, & 0.5938 \, & 0.5914 \, & 0.5871\,  & 0.5839 \, & 0.5836\,  & 0.5819\,  & 0.5806\\\hline\hline
%			\end{tabular}
%		\end{table}

%    	\begin{table}[h]
%		\centering
%		\caption{The measurement directions and the corresponding LHS bound for the $N$-setting linear EPR-steering inequalities given in \cite{NP2010}, here $N=2, 3, 4, 6, 10$.}
%		\begin{tabular}{cccccccc}
%			\hline\hline
%			n & 2      & 3      & 4      & 5     & 6      & 7      & 8      \\\hline
%			s & 0.7072 & 0.6667 & 0.6667 & /     & 0.6243 & /      & /      \\
%			o & 0.7072 & 0.6667 & 0.6319 & 0.618 & 0.6047 & 0.6033 & 0.5972 \\
%			\hline\hline
%			n & 9      & 10     & 12     & 14    & 16     & 18     & 20     \\\hline
%			s & /      & 0.6029 & /      & /     & /      & /      & /      \\
%			o & 0.5938 & 0.5914 & 0.5871 &0.5839 &0.5836  &0.5819  &0.5806 \\\hline\hline
%		\end{tabular}
%	\end{table}

\begin{figure}[h]
		\centering
\includegraphics[width=80mm]{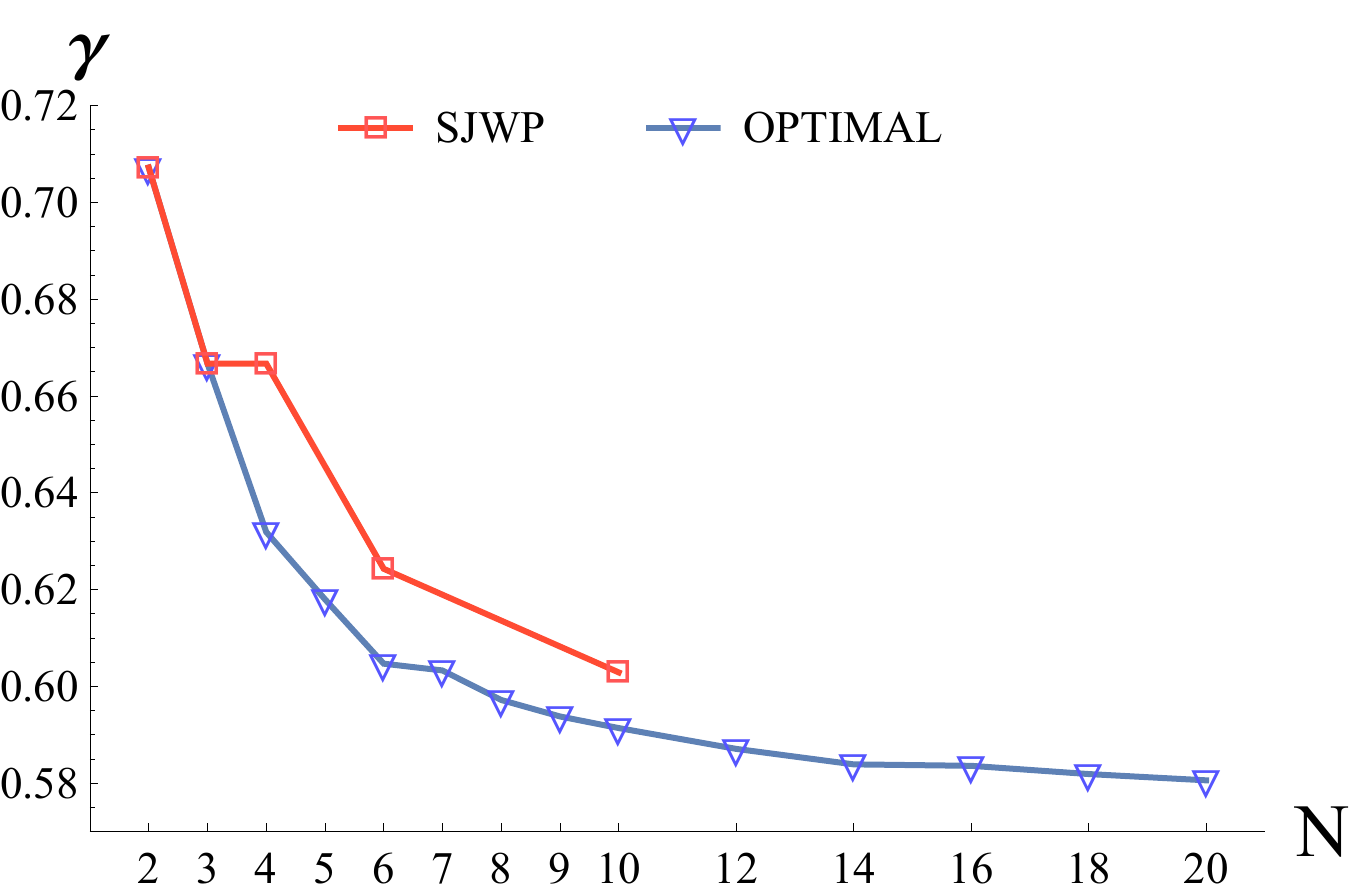}\hspace{3mm}
		\caption{Quantum violations of SJWP's inequalities and optimal inequalities for MEMS. The latter has a lower critical value $\gamma$ than the former.}\label{fig:MEMS}
\end{figure}

In Fig. \ref{fig:MEMS}, we have compared quantum violations of SJWP's inequalities and optimal inequalities for maximal entangled mixed state. We find that the latter has a lower critical value $\gamma$.

	\subsection{All-Versus-Nothing State}

The all-versus-nothing (AVN) state is given by~\cite{AVN2013}
\begin{eqnarray}
\rho_{\rm AVN}=\left[ \begin{array}{cccc}
             V\cos^2\theta & 0 & 0 & V\sin\theta\cos\theta    \\
             0 & (1-V)\sin^2\theta & (1-V)\sin\theta\cos\theta & 0 \\
             0 & (1-V)\sin\theta\cos\theta & (1-V)\cos^2\theta & 0 \\
            V\sin\theta\cos\theta & 0 & 0 &  V\sin^2\theta
           \end{array}\right],\label{state7}
\end{eqnarray}
with $\theta\in(0,\pi/2),\;V\in[0,1/2)\cup(1/2,1]$. By using the approach of all-versus-nothing proof, one can shown that such a state is always steerable~\cite{AVN2013}.

\begin{figure}[h]
		\centering
\includegraphics[width=80mm]{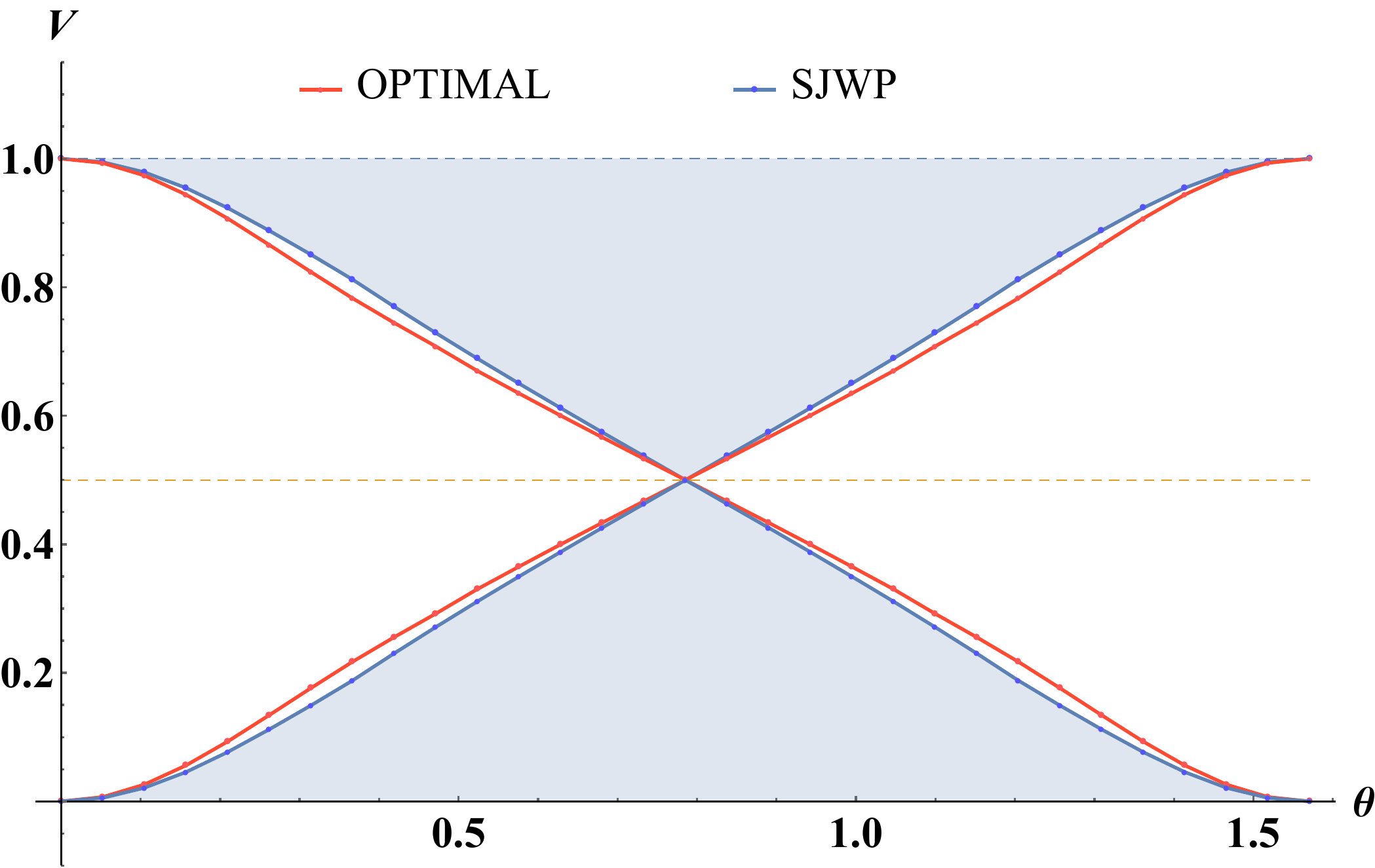}\hspace{3mm}
\includegraphics[width=80mm]{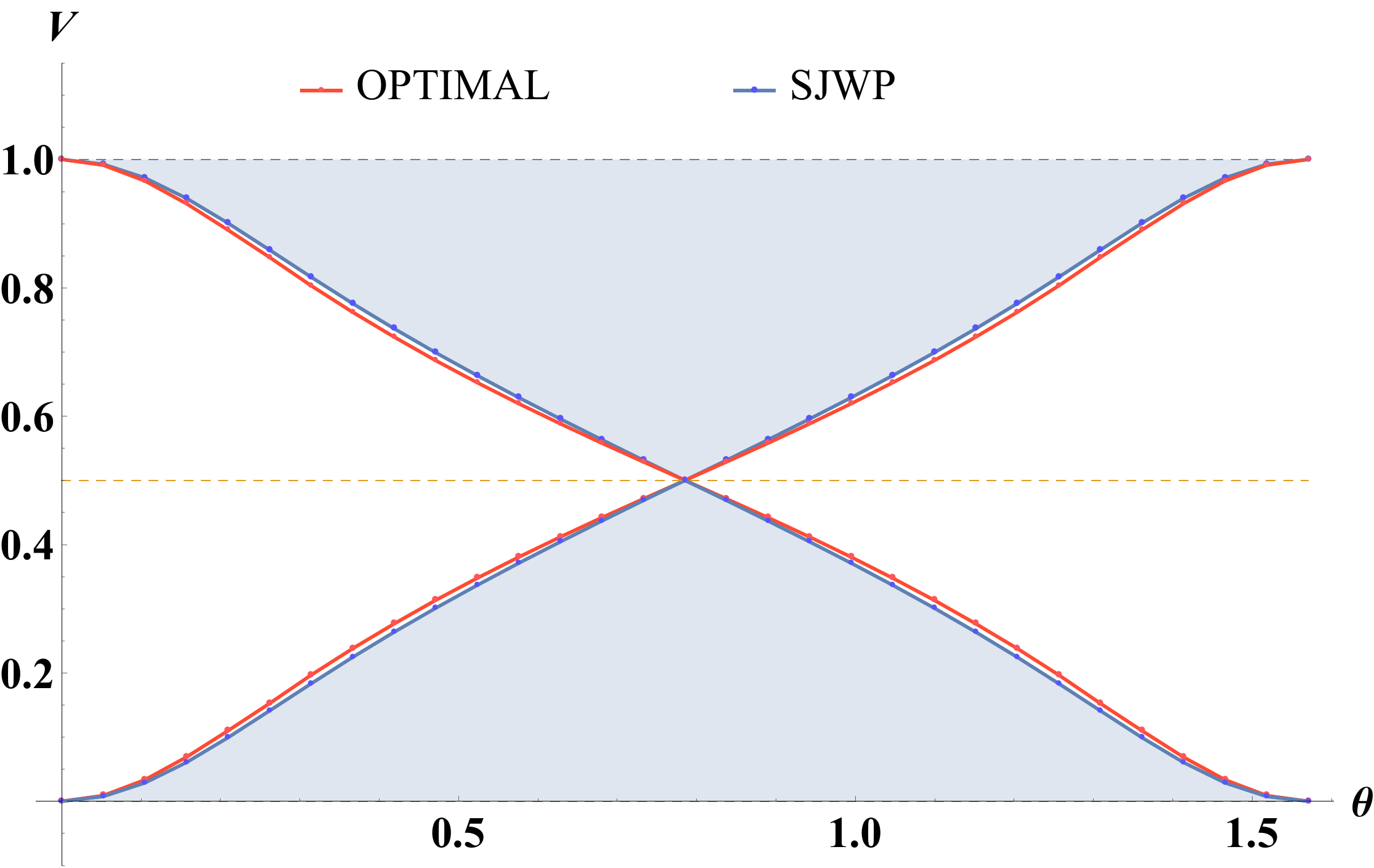}
		\caption{Quantum violations of SJWP's inequalities and optimal inequalities for AVN state (left: 4-setting; right: 6-setting). The latter can detect more regions of $(V, \theta)$ than the former.}\label{fig:AVN}
\end{figure}

In Fig. \ref{fig:AVN}, we have compared quantum violations of SJWP's inequalities and optimal inequalities for AVN state. We find that the detection ability of the latter is stronger than that of the former.

%\newpage
	
		\appendix